\documentclass[aps,twocolumn,superscriptaddress]{revtex4}
\usepackage{graphicx}   % for figures
\usepackage{epstopdf}
\usepackage{epsfig}
\usepackage{dcolumn}
\usepackage{bm}	
\usepackage{amsmath}			% Mode mathématique
\usepackage{amsfonts}			% Polices mathématiques
\usepackage{amssymb}			% Symboles mathématiques
\usepackage{latexsym}			% Symboles mathématiques avancés

\begin{document}

\title{Electromagnetic fluctuations and normal modes of a drifting relativistic plasma}
\author{C. Ruyer}\email{charles.ruyer@cea.fr}
\affiliation{CEA, DAM, DIF, F-91297 Arpajon, France}

\author{L. Gremillet}\email{laurent.gremillet@cea.fr}
\affiliation{CEA, DAM, DIF, F-91297 Arpajon, France}

\author{D. B\'enisti}
\affiliation{CEA, DAM, DIF, F-91297 Arpajon, France}

\author{G. Bonnaud}
\affiliation{CEA, Saclay, INSTN, F-91191 Gif-sur-Yvette, France}
\begin{abstract}
We present an exact calculation of the power spectrum of the electromagnetic fluctuations in a relativistic equilibrium plasma described by
Maxwell-J\"uttner distribution functions. We consider the cases of wave vectors parallel or normal to the plasma mean velocity. The relative
contributions of the subluminal and supraluminal fluctuations are evaluated. Analytical expressions of the spatial fluctuation spectra
are derived in each case. These theoretical results are compared to particle-in-cell simulations, showing a good reproduction of the
subluminal fluctuation spectra. 
\end{abstract}

\maketitle

\section{Introduction} 

An equilibrium, or near-equilibrium, plasma sustains a finite level of electromagnetic energy due to the random motion of particles
combined with the collective behaviour of the plasma. These spontaneous electromagnetic fluctuations play a major role in plasma
physics since they act as seeds for the instabilities driven by an input of free energy. Since Rostoker and Rosenbluth's pioneering
dressed-particle theory \cite[]{Rostoker_1960}, the fluctuation spectrum of a thermal plasma has been analyzed within several
frameworks in many papers and textbooks \cite[]{Sitenko_fluctuations, Akhiezer_fluctuations,Klimontovich_1982,Ichimaru_fluctuations,Tajima_1992,
Lund_1995,Opher_1996,Yoon_2007,Tautz_2007}. However, except for a few works restricted to isotropic particle distributions or
electrostatic modes \cite[]{Lerche_1968b,Stewart_1973, Klimontovich_1982}, all these studies were carried out in the nonrelativistic regime, and thus
did not address the spectrum induced by a plasma beam of relativistic temperature and mean drift velocity. Such a configuration is
of particular interest for the modeling of relativistic beam-plasma instabilities \cite[]{Bret_Gremillet_2010,Cottrill_2008} and
the related generation of collisionless shocks \cite[]{Lyubarsky_2006, Spitkovsky_2008b,Lemoine_2011b,Bret_2013}.  Schlickeiser, Lazar, Yoon and Felten
 \cite{Schlickeiser_Yoon_2012, Felten_Schlickeiser_2013, Lazar_Yoon_Schlickeiser_2013, Felten_Schlickeiser_2013a, Felten_Schlickeiser_2013b} 
 recently worked out a general fluctuation theory valid for unstable relativistic plasmas,
yet they performed practical calculations in the isotropic case only.

The objective of this paper is to present exact analytical and numerical calculations of the fluctuation spectra associated to
drifting relativistic plasmas described by a Maxwell-J\"uttner distribution function \cite[]{Wright_1975,Bret_Gremillet_Benisti_2010}.
We will  consider electromagnetic fluctuations propagating along or normal to the mean plasma velocity with different polarizations.
In each case, we will compute both the $(\omega,k)$-  and $k$-resolved spectra. To this goal, we will distinguish between the
contributions of the damped subluminal ($\omega/k<c$) and undamped supraluminal ($\omega/k>c$) normal modes. The latter will be computed from the electromagnetic
dispersion relation by generalizing the numerical scheme originally proposed by Fried and Gould \cite{Fried_Gould_1961} in the
nonrelativistic electrostatic regime. With a view on the Weibel-like filamentation instability of a relativistic plasma \cite[]{Yoon_2005,Achterberg_2007},
we will evaluate the spectrum of magnetic modes with wave vectors normal to the plasma drift velocity.

This paper is organized as follows. In Sec. \ref{sec:theory}, we recall the formalism of the standard fluctuation theory and adapt it to
Maxwell-J\"uttner distributions.  The spectrum of longitudinal and transverse electric fluctuations propagating along the plasma drift
velocity is treated in Sec. \ref{sec:parallel_fluc}, whereas the spectrum of magnetic fluctuations propagating normal to the plasma drift
velocity  is computed in Sec. \ref{sec:mag_fluc}. Section \ref{sec:PIC_sims} confronts our theoretical formula to particle-in-cell (PIC)
simulations. Finally, our results are summarized in Sec. \ref{sec:conclusions}

\section{Electromagnetic fluctuation theory}\label{sec:theory}

\subsection{General formalism}\label{subsec:formalism}

Let us consider a uniform relativistic plasma composed of a number of charged particle species of mass $m_s$, charge $q_s$ and density $n_s$.
According to Refs. \cite[]{Sitenko_fluctuations, Ichimaru_fluctuations}, assuming an adiabatic switch-on of the electromagnetic interactions,
the spectral density tensor of the plasma electric fluctuations writes
\begin{equation}\label{eq:fluctuations}
  \langle \mathbf{EE}^{\dagger}\rangle_{\mathbf{k},\omega}=\mathbf{Z}_{\mathbf{k},\omega} \cdot 
  \langle \mathbf{jj^{\dagger}}\rangle_{\mathbf{k},\omega}\cdot \mathbf{Z}^{\dagger}_{\mathbf{k},\omega} \, ,
\end{equation}
where the fluctuation source $\langle \mathbf{jj^{\dagger}}\rangle_{\mathbf{k},\omega}$ is the spectral density tensor of the ballistic plasma
current density 
\begin{equation}\label{eq:source1}
  \langle j_\alpha j_\beta^* \rangle_{\mathbf{k},\omega} =2\pi\epsilon_0\sum_s m_s\omega_{ps}^2 \int_{\mathbb{R}^3} d^3p\, v_\alpha v_\beta f_s^{(0)}(\mathbf{p}) 
  \delta(\omega-\mathbf{k}\cdot \mathbf{v}) \,,
\end{equation}
with $f_s^{(0)}(\mathbf{p})$ the equilibrium distribution function and $\omega_{ps}^2 = n_s q_s^2/m_s\epsilon_0$ the plasma frequency of the $s$th
species. The resistivity tensor $\mathbf{Z}_{\mathbf{k},\omega}$ is defined from the linear relation
\begin{equation}
  \mathbf{E}_{\mathbf{k},\omega}=\mathbf{Z}_{\mathbf{k},\omega} \cdot \mathbf{j}_{\mathbf{k},\omega} \, .
\end{equation}
We adopt the following conventions for the Fourier transform of a given function $g(\mathbf{r},t)$ 
\begin{align}
  &g_{\mathbf{k},\omega}= \iiiint_{\mathbb{R}^4} d^3r dt\, g(\mathbf{r},t) \exp{\left[i(\omega t - \mathbf{k} \cdot \mathbf{r})\right]} \, , \\
  &g(\mathbf{r},t)= \iiiint_{\mathbb{R}^4} \frac{d^3k d\omega}{(2\pi)^4}\,g_{\mathbf{k},\omega} \exp{\left[i(\mathbf{k} \cdot \mathbf{r} - \omega t)\right]} \, .
\end{align}

For a plasma described by gyrotropic equilibrium distribution functions $f_s^{(0)}(p_\perp^2,p_z)$, the wave vector of
the fluctuations can be chosen in the $(k_y,k_z)$ plane without loss of generality. The resistivity tensor $\mathbf{Z}_{\mathbf{k},\omega}$ then reads 
\begin{widetext}
\begin{equation} \label{eq:Z}
\mathbf{Z}_{\mathbf{k},\omega} = -i\frac{\omega}{\epsilon_0} \left( \begin{array}{ccc}
 \frac{1}{\omega^2 \epsilon_{xx}-k^2c^2}& 0  & 0 \\
 0 & \frac{\omega^2 \epsilon_{zz}-k_y^2c^2}{(\omega^2 \epsilon_{zz}-k_y^2c^2)(\omega^2 \epsilon_{yy}-k_z^2c^2)-(\omega^2 \epsilon_{zy}+k_yk_z c^2)^2}& 
 	-\frac{\omega^2 \epsilon_{zy}+k_yk_zc^2}{(\omega^2 \epsilon_{zz}-k_y^2c^2)(\omega^2 \epsilon_{yy}-k_z^2c^2)-(\omega^2 \epsilon_{zy}+k_yk_z c^2)^2} \\
 0 & -\frac{\omega^2 \epsilon_{zy}+k_yk_z c^2}{(\omega^2 \epsilon_{zz}-k_y^2c^2)(\omega^2 \epsilon_{yy}-k_z^2c^2)-(\omega^2 \epsilon_{zy}+k_yk_zc^2 )^2} &
  	\frac{\omega^2 \epsilon_{yy}-k_z^2c^2}{(\omega^2 \epsilon_{zz}-k_y^2c^2)(\omega^2 \epsilon_{yy}-k_z^2c^2)-(\omega^2 \epsilon_{zy}+k_yk_zc^2 )^2}        
 \end{array}\right).
\end{equation}
\end{widetext}
We have introduced the dielectric tensor elements \cite[]{Ichimaru_1973}
\begin{align}\label{eq:tenseur_eps}
\boldsymbol{\epsilon}_{\alpha \beta}(\mathbf{k},\omega) &= \delta_{\alpha \beta} 
+\sum_s \frac{\omega_{ps}^2}{\omega^2} \iiint d^3p \,   \frac{p_{\alpha}}{\gamma} \frac{\partial f_s^{(0)}}{\partial p_{\beta}} \nonumber \\
&+ \sum_s \frac{\omega_{ps}^2}{\omega^2} \iiint d^3p \,   v_\alpha \frac{p_{\beta}}{\gamma}
\frac{\mathbf{k}\cdot \partial f_s^{(0)}/ \partial \mathbf{p}}{\omega-\mathbf{k}\cdot \mathbf{v}}  \, .
\end{align}
The equation $\vert \mathbf{Z}_{\mathbf{k},\omega} \vert = 0$ defines the dispersion relation of the discrete normal modes of the system 
$\omega_l(\mathbf{k})$:
\begin{align}
  &\omega^2 \epsilon_{xx}-k^2c^2 = 0 \, , \label{eq:dispe1}\\
  &(\omega^2 \epsilon_{zz}-k_y^2c^2)(\omega^2 \epsilon_{yy}-k_z^2c^2)-(\omega^2 \epsilon_{zy}+k_yk_zc^2 )^2 = 0 \,.\label{eq:dispe2}
\end{align}

\subsection{Maxwell-J\"uttner distribution functions} \label{subsec:mj}

From now on, the particle species are assumed to obey drifting Maxwell-J\"uttner distribution functions \cite[]{Wright_1975}: 
\begin{equation}\label{eq:mj}
  f_s^{(0)}(\textbf{p}) = F_s\exp{[-\mu_s( \gamma - \beta_{ds}p_z )] } \, ,
\end{equation}
where $\gamma = \sqrt{1+p^2/(m_sc)^2}$ is the relativistic factor, $\mu_s=m_sc^2/T_s$ the normalized inverse temperature and $\beta_{ds}=\langle v_z/c \rangle$
the $z$-aligned mean velocity.  We have introduced the normalization factor $F_s =\mu_s/4\pi (m_sc)^3\gamma_{ds}^{2} K_2(\mu_s/\gamma_{ds})$,
with $\gamma_{ds}=(1-\beta_{ds}^2)^{-1/2}$ and $K_2$ the modified Bessel function of the second kind. The distribution functions are normalized
to unity: $\int f_s^{(0)}(\mathbf{p})d^3p=1$. Charge and current neutralization is assumed so as to ensure a field-free equilibrium. 

Let us introduce $\alpha$, the angle between the wave vector $\mathbf{k}$ and the $z$-axis, and $\beta_\phi=\omega/kc$, the normalized wave phase
velocity.  By changing to velocity variables in cylindrical coordinates along $\mathbf{k}$, the triple integrals involved in the dielectric tensor
can be reduced to the following one-dimensional quadratures \cite[]{Bret_Gremillet_2010}:
\begin{align}
  \epsilon_{xx}&=1-2\pi \sum_{s}\frac{F_s \mu_{s} \omega_{ps}^2}{\omega^2} \left(\beta_{\phi}-\beta_{ds} \cos \alpha \right) D_s \label{eq:esilonxx}  \, ,\\
  \epsilon_{yy}&=1-2\pi \sum_{s}\frac{F_s \mu_{s}\omega_{ps}^2}{\omega^2} \left(\beta_{\phi}-\beta_{ds} \cos \alpha  \right) \nonumber \\
  &\times \left( A_s\cos^2 \alpha + 2C_s \cos \alpha \sin \alpha + B_s \sin^2 \alpha \right) \label{eq:esilonyy}   \, ,\\
  \epsilon_{zz}&=1-2\pi \sum_{s}\frac{ F_s \mu_{s}\omega_{ps}^2}{\omega^2} \left(\beta_{\phi}-\beta_{ds} \cos \alpha  \right)  \nonumber \\
  &\times \left( A_s \sin^2 \alpha - 2C_s \cos \alpha \sin \alpha + B_s \cos^2 \alpha \right)   \nonumber \\
  &+\sum_{s}\frac{\mu_{s}\omega_{ps}^2}{\omega^2}\beta^2_{ds} \label{eq:esilonzz}  \, ,\\
  \epsilon_{yz}&= - 2\pi \sum_{s}\frac{F_s \mu_{s}\omega_{ps}^2}{\omega^2} \left(\beta_{\phi}-\beta_{ds} \cos \alpha  \right) \nonumber \\
  &\times \left[ (B_s-A_s) \cos \alpha \sin \alpha + C_s(\cos^2 \alpha-\sin^2 \alpha ) \right] \label{eq:esilonyz} \, .
\end{align}
Each function $X \in \{A_s, B_s, C_s, D_s\}$ is defined as
\begin{equation} \label{eq:X1}
  X(\beta_{\phi}) = \int_{-1}^{1}d\beta \, \frac{f_{X}(\gamma,\rho_s,\nu_s)}{\beta_\phi-\beta} \quad (\Im \beta_\phi > 0)  \, ,
\end{equation}
with 
\begin{align}\label{eq:A-B-C-D}
  &f_{A_s}= \frac{ \gamma e^{-h_s} }{h_s^5} \left[ (h_s+1) (\rho_s^2+2\nu_s^2) + \nu_s^2 h_s^2\right] \nonumber \, , \\
  &f_{B_s} = \frac{ \beta^2 \gamma^3 e^{-h_s}}{h_s^5} \left[ (h_s+1) (2\rho_s^2+\nu_s^2) + \rho_s^2 h_s^2 \right] \nonumber \, , \\
  &f_{C_s} = -\nu_s \frac{ \beta \gamma^2 \rho_s e^{-h_s} }{h_s^5} \left[ 3(h_s+1 ) + h_s^2 \right] \, , \\
  &f_{D_s} =\frac{ \gamma e^{-h_s} }{h_s^{3}}(h_s+1) \,, \nonumber
\end{align}
where
\begin{align}
  &\rho_s = \mu_s \gamma (1-\beta_{ds}\beta \cos \alpha) \,, \label{eq:rhos} \\
  &\nu_s = \mu_s \beta_{ds} \sin \alpha \,, \label{eq:nus} \\
  &h_s=(\rho_s^2-\nu_s^2)^{1/2} \, . \label{eq:hs}
\end{align}

Making use of the above definitions, the source tensor \eqref{eq:source1} simplifies to 
\begin{align}
&\langle \mathbf{jj}^{\dagger}_{xx}\rangle_{\mathbf{k},\omega}=(2\pi)^2 \epsilon_0 H(1-|\beta_\phi|) \sum_s \frac{F_s m_s\omega_{ps}^2}{k}
 \left[ f_{D_s} \right]_{\beta=\beta_\phi} \, , \label{eq:sourcexx} \\
&\langle \mathbf{jj}^{\dagger}_{yy}\rangle_{\mathbf{k},\omega}=(2\pi)^2\epsilon_0 H(1-|\beta_\phi|) \sum_s \frac{ F_s m_s\omega_{ps}^2}{k} \nonumber \\
&\times \left[ f_{A_s}\cos^2\alpha + f_{B_s} \sin^2\alpha  + 2f_{C_s} \cos \alpha \sin \alpha \right]_{\beta=\beta_\phi} \, , \label{eq:sourceyy}\\
&\langle \mathbf{jj}^{\dagger}_{zz}\rangle_{\mathbf{k},\omega}=(2\pi)^2 \epsilon_0 H(1-|\beta_\phi|) \sum_s \frac{m_s F_s \omega_{ps}^2}{k} \nonumber \\
&\times \left[f_{A_s}\sin^2 \alpha + f_{B_s}\cos^2 \alpha- 2f_{C_s}\cos \alpha \sin \alpha \right] _{\beta=\beta_\phi} \, , \label{sourcezz}\\
&\langle \mathbf{jj}^{\dagger}_{yz}\rangle_{\mathbf{k},\omega}=(2\pi)^2 \epsilon_0 H(1-|\beta_\phi|) \sum_s \frac{m_s F_s \omega_{ps}^2}{k} \nonumber \\
&\times \left[(f_{B_s} - f_{A_s} )\cos \alpha \sin \alpha + f_{C_s}(\cos^2 \alpha -\sin^2 \alpha) \right]_{\beta=\beta_{\phi}} \, , \label{eq:sourceyz}
\end{align}
where $H(x)$ denotes the step function. 

\subsection{Branch cuts}

In order to compute Eq. \eqref{eq:fluctuations} for $(\omega,k) \in \mathbb{R}^2$, the function $X \in \{A,B,C,D\}$ (the index $s$ is omitted for
the sake of clarity) must be analytically continued to the $\Im \beta_\phi \le 0$ complex half-plane. To this goal, it is convenient to use the
following expression
\begin{align}\label{eq:bc}
  X(\beta_\phi) =& -\int_{-1}^{1} d\beta\, \frac{f_X(\beta)-f_X(\beta_{\phi})}{\beta-\beta_{\phi}}   \nonumber \\
   &- f_X(\beta_{\phi}) \ln \left(\frac{\beta_{\phi}-1}{\beta_{\phi}+1}\right)\, ,
\end{align}
where $\ln$ denotes a particular branch of the complex logarithm to be defined. The integrand of Eq. \eqref{eq:bc} has no singularity, and
therefore allows a standard numerical integration scheme. Because $X$ has to be computed on the real $\beta_\phi$ axis, the logarithm's
branch cuts are chosen to be $]-1-i\infty,-1[\cup]+1,+1-i\infty[$. This implies that the phase angles $\theta_{\beta\pm 1}=\mathrm{arg} (\beta \pm 1)$
lie within the interval $-\pi/2<\theta_{\beta \pm 1}< 3\pi/2$. This specification leads to the same branch cuts for the multivalued relativistic
factor
\begin{equation}\label{gamma}
  \gamma = i(|\beta-1||\beta+1|)^{-1/2}\exp \left[-i(\theta_{\beta+1} + \theta_{\beta-1})\right] \, .
\end{equation}

Similarly, the analytic continuation of the function $h(\beta)$ \eqref{eq:hs} follows from its factorized form
\begin{flalign}\label{eq:hs2}
&h(\beta)=\mu \gamma \sqrt{(\beta_d\beta-\zeta_+)(\beta_d \beta-\zeta_-)}  \, , \\
&\zeta_\pm = \cos \alpha \pm i\frac{\sin \alpha }{\gamma_d} \, .
\end{flalign}
The phase angles $\theta_{\beta_d \beta-\zeta_{\pm}} = \mathrm{arg}(\beta_d \beta-\zeta_{\pm})$ are now restricted to the intervals
$-\pi/2<\theta_{\beta_d \beta- \zeta_-}< 3\pi/2$ and $-3\pi/2<\theta_{\beta_d\beta - \zeta_+}< \pi/2$, which leads to the branch cuts
$]-1-i\infty,-1[\cup]+1,+1-i\infty[\cup]\zeta_{-} -i\infty,\zeta_-[\cup]\zeta_{+}+i\infty,\zeta_+[$.  Equation (\ref{eq:X1}) shows
that the function $X$ inherits the branch cuts $]-1-i\infty,-1[\cup]\zeta_{-} -i\infty,\zeta_-[$ in the lower half $\beta_\phi$ plane.
By contrast, it is of the Cauchy type in the upper half $\beta_\phi$ plane, and therefore everywhere holomorphic.

\begin{figure}
\centering
\includegraphics[scale=1]{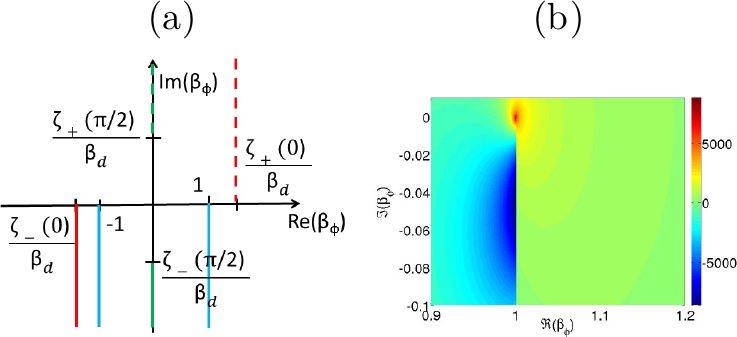}
\caption{ \label{fig:map_B}(a) Branch cuts (colored solid lines) of the function $X \in \{A,B,C,D\}$ in the complex $\beta_\phi$ plane for $\alpha = 0$
and $\alpha = \pi/2$. (b) Map of $\Re B$ in the complex $\beta_\phi$ plane for $\mu=1$, $\beta_d=0.9$ and $\alpha=0$.}
\end{figure}

In the following, we will compute the $(\omega,k)$-resolved spectra of the longitudinal and transverse fluctuations propagating along the beam
direction ($\alpha=0$), and of the magnetic fluctuations propagating normal to the beam direction ($\alpha = \pi/2$).

\section{Fluctuations with wave vectors parallel to the plasma drift velocity} \label{sec:parallel_fluc}

\subsection{Longitudinal fluctuations}\label{section:ez2}

\subsubsection{Basic formulae}

Here, we consider electrostatic fluctuations with $\mathbf{E} \parallel \mathbf{k}$ and $\mathbf{k} \parallel \mathbf{z}$. Combining
Eqs. \eqref{eq:fluctuations} and \eqref{eq:Z} yields
\begin{equation}
  \langle E_z E_z^*\rangle_{k_z,\omega} = \frac{\langle j_zj_z^*\rangle_{k_z,\omega}}{\epsilon_0^2 \omega^2 |\epsilon_{zz}|^2} \,, \label{eq:fluctuzz}
\end{equation}
where $\epsilon_{zz}$ can be rewritten as (see \ref{part:exact_dispe_zz})
\begin{equation}  \label{eq:epszz}
\epsilon_{zz} = 1-\sum_{s}\frac{2\pi F_s \mu_{s}\omega_{ps}^2}{k_z^2c^2} (\beta_{\phi}-\beta_{ds}) \tilde{B}_s(\beta_\phi) 
+ \sum_{s}\frac{\mu_{s}\omega_{ps}^2}{k_z^2c^2} \, , 
\end{equation}
with $\tilde{B}_s(\beta_\phi)=\int d\beta \, f_B/\beta^2(\beta_\phi-\beta)$. Upon defining the susceptibilities $\chi_s$ from the standard relation
$\epsilon_{zz} = 1+\sum_s \chi_s$, and noting that $\langle j_zj_z^*\rangle_s$ is proportional to the singularity $f_{B_s}(\beta_\phi)$ of the integrand defined in $B_s$, the electric
fluctuation spectrum can be further simplified as
\begin{equation} \label{eq:fluctuzz2}
\langle E_z E_z^*\rangle_{k_z,\omega} = \frac{2}{\epsilon_0|\epsilon_{zz}|^2}  \sum_s \frac{T_s \Im (\chi_s)}{\omega - k_z v_{ds}} \, ,
\end{equation}
This expression generalizes the fluctuation-dissipation theorem \cite{Akhiezer_fluctuations} to the case of multiple Maxwell-J\"uttner-distributed
particle species of various drift velocities and temperatures. Note that it is formally identical to the formula derived for nonrelativistic
drifting Maxwellians by Lund \emph{et al.} \cite{Lund_1995}.

\subsubsection{Dispersion relation}\label{part:exact_dispe_zz}

The fluctuation spectrum [Eq. \eqref{eq:fluctuzz2}] is strongly peaked around the weakly-damped and undamped solutions of $\epsilon_{zz}(\omega,k_z)=0$.
Because it implicitly assumes unbounded particle velocities, the standard nonrelativistic kinetic description of a stable plasma wrongly predicts that
all of its eigenmodes are Landau-damped, whatever their phase velocity. Now, it can be proved from a more rigorous relativistic description that only
the subluminal modes (with $\omega/k < c$) are damped \cite{Lerche_1969, Schlickeiser_2004}. For the sake of numerical convenience, we recast the
dispersion relation in the form
\begin{align}\label{eq:supralumzz}
  k_z^2c^2 &= \sum_s 2\pi F_s \mu_s \omega_{ps}^2 (\beta_{\phi}-\beta_{ds})  \tilde{B}_s(\beta_\phi) -\sum_{s}\mu_{s}\omega_{ps}^2 \nonumber \\
  &= \mathcal{G}(\beta_\phi) \,.
\end{align}
This formulation, in which $k_z^2 (>0)$ is a function of $\beta_\phi$ only, lends itself to the efficient numerical scheme introduced by Fried and
Gould \cite{Fried_Gould_1961} in a nonrelativistic framework. This technique consists, first, in determining the locus of the zeroes of $\Im \mathcal{G}(\beta_\phi)$.
This can be readily performed by means of a contour plot in a finely discretized portion of the complex $\beta_\phi$ plane. Then, we retain those zeroes
fulfilling $\Re \mathcal{G}(\beta_\phi) > 0$ and identify $k_z c=\sqrt{\Re \mathcal{G}(\beta_\phi)}$. Depending on the $\beta_\phi$ domain considered, this
method allows us to simultaneously solve for a set of discrete electrostatic ($\mathbf{E} \parallel \mathbf{k}$) solutions $\omega_L(k_z)$.

As shown in Refs. \cite{Lerche_1969,Laing_Diver_2006} for an isotropic plasma ($\beta_{ds}=0$), the supraluminal electrostatic modes (with $\omega/k > c$) exist only
over a finite interval $0  \le k_z \le k_{c\pm}$ (for a positive wavenumber). The critical value $k_{c\pm}$ defined by $\omega_L(k_{c\pm}) = \pm k_{c\pm} c$, depends on the sign of the
phase velocity. In the general case, $k_{c\pm}$ can be obtained by setting $\beta_\phi = \pm 1$ in Eq. \eqref{eq:supralumzz}, yielding
\cite[]{McKee_1971, Laing_Diver_2006}
\begin{equation}\label{eq:kc}
  k_{c\pm}^2 c^2 = \sum_s (1\pm\beta_{ds})^2\gamma_{ds}\omega_{ps}^2
  \frac{K_1(\frac{\mu_s}{\gamma_{ds}}) + 2\frac{\gamma_{ds}}{\mu_s}K_0(\frac{\mu_s}{\gamma_{ds}})}{K_2(\frac{\mu_s}{\gamma_{ds}})} \,.
\end{equation}
Consequently, for a fixed sign of the phase velocity, there is a maximum of one supraluminal longitudinal mode ($\omega_{LS+}$ or $\omega_{LS-}$).

\subsubsection{$(\omega,k)$-resolved spectrum}

\begin{figure*}
\centering
\includegraphics[scale=1]{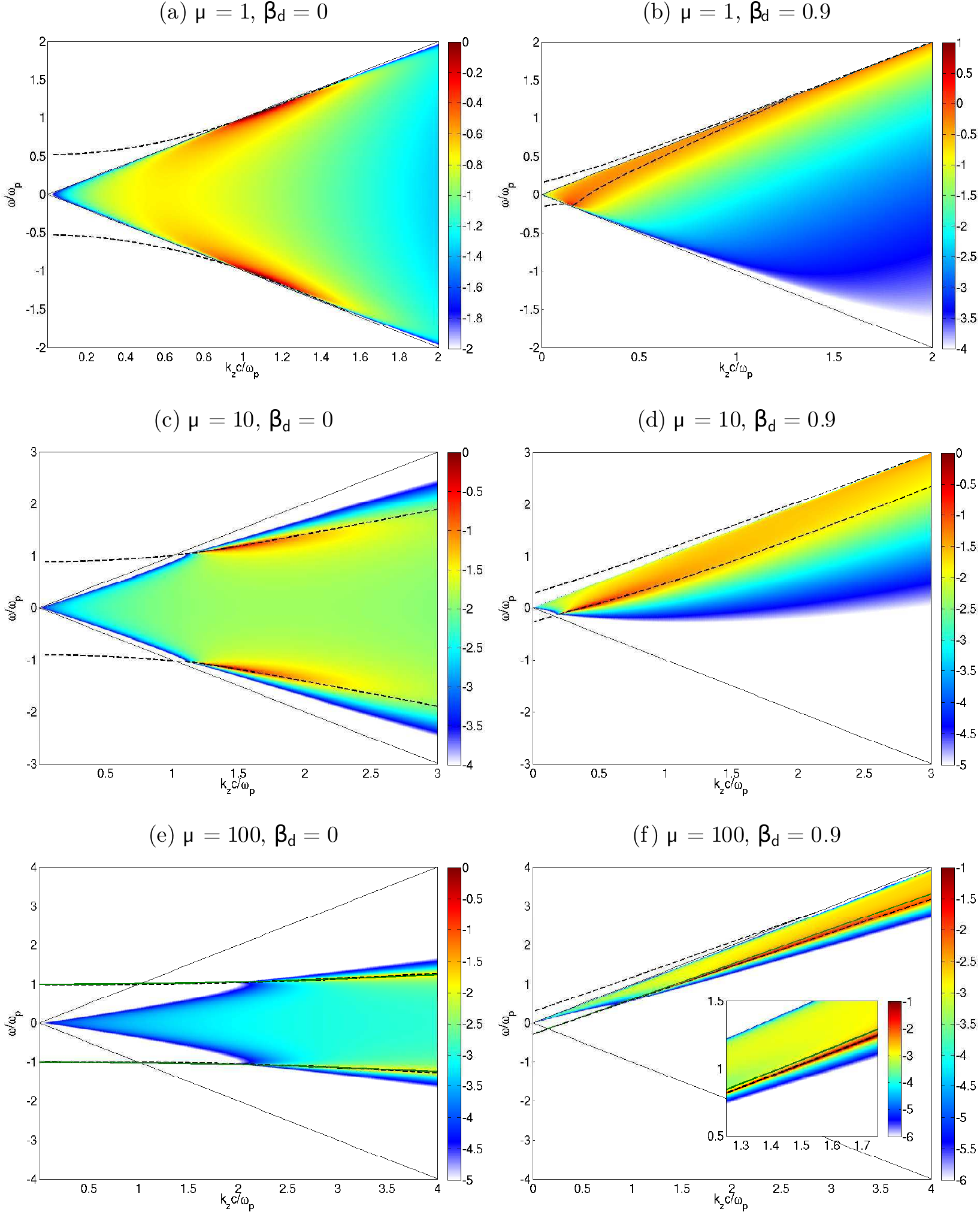}
\caption{\label{ez2_spectrum_alpha0}Power spectrum $\langle E_zE_z^*\rangle_{k_z,\omega}$ (normalized to $m_e^2c^3/e^2$) in $\log_{10}$ scale
for an $e^-e^+$ pair plasma and various $\mu$ and $\beta_d$ values. The two black solid lines delimit the subluminal region ($\vert  \beta_{\phi} \vert \le 1 $).
The black dashed curves plot the exact eigenmodes computed from Eq. \eqref{eq:supralumzz}. In the low-temperature case ($\mu=100$),
the green solid line plots the Bohm-Gross mode in panel (e) and the Lorentz-transformed Bohm-Gross mode in panel (f). }
\end{figure*}

Figures \ref{ez2_spectrum_alpha0}(a-f) display the longitudinal fluctuation spectra of a pair plasma (or, equivalently, of an electron
plasma with a neutralizing background) of various drift velocities and temperatures (equal for all species). The main difference between
these results and previously published nonrelativistic calculations \cite{Lund_1995,Tautz_2007,Yoon_2007,Schlickeiser_Yoon_2012} is the
cutoff occurring for supraluminal modes. Since the quadratures defined by Eq. \eqref{eq:X1} have no pole for $\vert \beta_\phi \vert > 1$,
the dielectric tensor is purely real; hence, supraluminal modes cannot be excited by an inverse-Landau (Cerenkov) type mechanism.
For a pair plasma, Eq. \eqref{eq:fluctuzz2} is proportional to  $\Im (\epsilon_{zz})/|\epsilon_{zz}|^2 $. As shown in Ref. \cite[]{Klimontovich_1982} 
this implies that $ \langle E_z E_z^*\rangle_{k_z,\omega} \propto  \delta(\epsilon_{zz})$ in the supraluminal region ($\vert  \beta_{\phi} \vert > 1 $). 
Consequently, the supraluminal part of the fluctuation spectra only 
results from the delta-like singularities (in the absence of collisions) associated to supraluminal eigenmodes, solution of Eq. \eqref{eq:supralumzz}.

As expected, the subluminal fluctuation spectra exhibit strong maxima along the weakly-damped part of the eigenmode curves. The latter
(plotted as black dashed curves) intersect the $\beta_\phi = \pm 1$ lines at $k_z = k_{c\pm}$. The fluctuation maxima are all the sharper
when the plasma temperature drops, being increasingly hard to capture numerically.
In the isotropic, low-temperature case ($\mu = 100$)
of Fig. \ref{ez2_spectrum_alpha0}(e), the well-known nonrelativistic Bohm-Gross mode $\omega = \omega_p(1+3k_z^2c^2/2\omega_p^2 \mu)$
is found to closely match the exact eigenmode.  
We checked that the fluctuation spectrum significantly broadens away from the eigenmode for $k_zc/\omega_p \gtrsim 3$,
which corresponds to the well-known strongly damped regime $k_z \lambda_D>0.3$.
In the isotropic, relativistically-hot case ($\mu =1$), the spectrum
broadens away from the eigenmode curves, attaining comparable values over most of the subluminal cone for $k_z c/\omega_p < 1.6$. 

The influence of a relativistic drift velocity ($\beta_d=0.9$) is illustrated in Figs. \ref{ez2_spectrum_alpha0}(b,d,f). Both the spectra
and the dominant eigenmodes turn asymmetric with respect to $\omega = 0$.  The fluctuations peak close to the eigenmode, although only the slowest is visible on Figs. \ref{ez2_spectrum_alpha0}(b,d,f). In the
low-temperature case ($\mu=100$), one can approximate the real frequency of the exact eigenmodes by Lorentz transformation of the Bohm-Gross
modes [Fig. \ref{ez2_spectrum_alpha0}(f)]. Consistently with Eq. \eqref{eq:kc}, the supraluminal eigenmodes are found only for a limited
range of wave vectors $k_z \in[-k_{c-},k_{c+}]$ with $k_{c-}\neq k_{c+}$. 

\subsubsection{$k$-resolved spectrum} \label{sec:long_k_spec}

\begin{figure}
\centering
\includegraphics[scale=1]{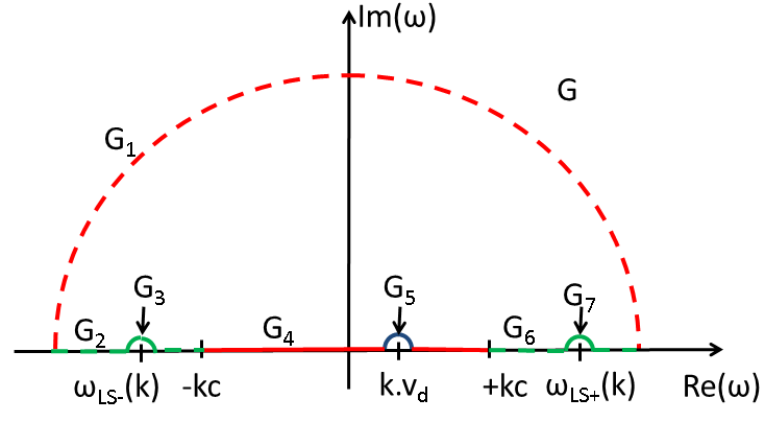}
\caption{\label{fig:contour_G} Closed contour $G=\cup_{i=1}^7 G_i$ in the complex $\omega$-plane used in Eq. \eqref{eq:contour_G}. The arrows indicate
the ballistic singularity $k\beta_d$ and the supraluminal eigenmodes $\omega_{LS\pm}(k)$.
}
\end{figure}

The spatial fluctuation spectrum is of interest both experimentally and theoretically. The integration of Eq. \eqref{eq:fluctuzz2} over $\omega$ gives
\begin{equation}\label{eq:intez2}
\langle E_z E_z^*\rangle_{k_z} = -\frac{2T}{\epsilon_0} \int_{-\infty}^{+\infty} \frac{d\omega}{2\pi} \frac{1}{(\omega - k_z v_{d})} \Im \left(\frac{1}{\epsilon_{zz}}\right)
\end{equation}
This integration can be carried out by the method introduced in Ref. \cite[]{Langdon_1979} in the non-relativistic case. To this effect, let us rewrite
 Eq. \eqref{eq:intez2} as
\begin{align}\label{eq:intez3}
  \langle E_z E_z^*\rangle_{k_z} &=- \frac{1}{\epsilon_0} \int_{-\infty}^{+\infty} \frac{d\omega}{2\pi} \Im{\left[\frac{2T}{(\omega - k_z v_{d})}
  \left( \frac{1}{\epsilon_{zz}}-1\right)\right]}  \nonumber \\
  &= \Im \int_{-\infty}^{+\infty} \frac{d\omega}{2\pi} I(\omega,k_z) \, ,
\end{align}
where we have defined
\begin{equation}\label{eq:I}
  I(\omega,k_z) =- \frac{2T}{\epsilon_0  (\omega - k_z v_{d})} \left(\frac{1}{\epsilon_{zz}}-1 \right) \,.
\end{equation}

\begin{figure}
\centering
\includegraphics[scale=1]{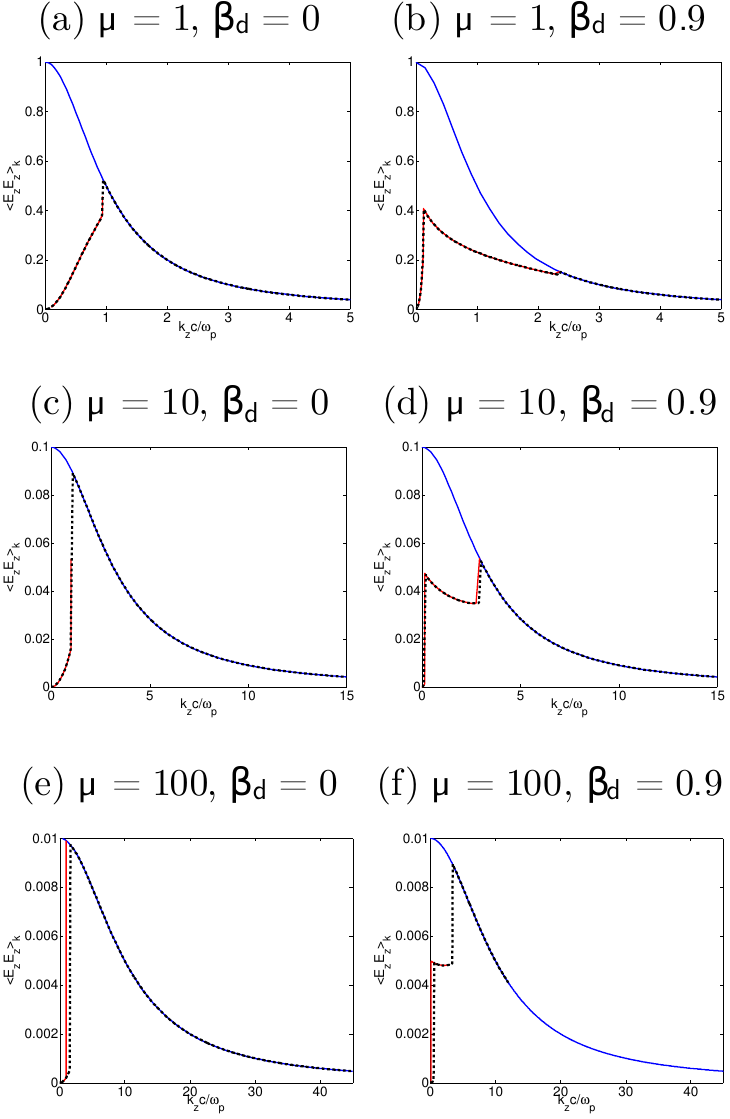}
\caption{\label{fig:intez2_alpha0} Spatial fluctuation spectra  $\langle E_zE_z^*\rangle_{k_z}$ (normalized to $\frac{ m_e^2c^3}{e^2}
(\sum_s \omega_{ps}^2)^{1/2}$) for an electron/pair plasma and various $\mu$ and $\beta_d$ values (identical for all species). The blue and red
lines correspond to the total and subluminal spectra computed from Eqs. \eqref{eq:int_ezz} and \eqref{eq:int_sublumzz}, respectively.
The black-dotted lines correspond to the numerical integration of Eq. (\ref{eq:fluctuzz2})
over the subluminal domain.}
\end{figure}
We now introduce the subluminal fluctuation spectrum $\langle E_i E_j^*\rangle^{|\beta_\phi|<1}_{k}$ which is the integration of $I(\omega,k)$ over the subluminal $\omega$-domain 
$[ -kc, +kc]$. The supraluminal fluctuation spectrum $\langle E_i E_j^*\rangle^{|\beta_\phi|>1}_{k}$ follows by the integration over 
$\omega \in [-\infty, -kc ] \cup [+kc,+\infty]$.

This function is analytic in the upper half $\omega$-plane but has several singularities on the real $\omega$-axis. The first one is due to the
ballistic term $(\omega-k v_d)^{-1}$ and is located in the subluminal region $|\omega/kc|<1$. The others correspond to the supraluminal (undamped)
eigenmodes $\omega_{LS\pm}(k)$. There follows
\begin{equation}\label{eq:contour_G}
  \sum_{i=1}^7 \Im \int_{G_i}d\omega I(\omega,k_z)=0 \,,
\end{equation}
where the closed integration contour $G=\cup_{i=1}^7 G_i$ is drawn in Fig. \ref{fig:contour_G}.  Since $\lim_{|\omega| \to +\infty} \left(\epsilon_{zz}^{-1}-1 \right) = 0$
and $I\in \mathbb{R}$ for $\vert \omega/kc \vert > 1$, the integrals over the $G_1$, $G_2$ and $G_6$ contours vanish, yielding
\begin{equation}\label{eq:intgzz}
\langle E_z E_z^*\rangle ^{|\beta_\phi|<1}_{k_z}  - \pi \sum_{\omega=\omega_{LS\pm}} \mathrm{Res} (I)_{\omega}= - \pi \mathrm{Res} (I)_{k_z v_d} \, .
\end{equation}
To obtain Eq. \eqref{eq:intgzz}, the $G_4$ term has been identified with the subluminal part of fluctuation spectrum, 
$\langle E_z E_z^*\rangle ^{\vert \beta_\phi \vert<1}_k$, and the integrals over the semi-circle contours $G_3$, $G_5$ and $G_7$ have been evaluated
using the residue theorem. The second term on the left hand side corresponds to the supraluminal fluctuation spectrum:
\begin{align} \label{eq:int_supralumzz}
\langle E_z E_z^*\rangle^{|\beta_\phi|>1}_{k_z} &=  - \pi\sum_{\omega=\omega_{LS\pm}}  \mathrm{Res}(I)_{\omega} \\ \nonumber
  &= \sum_{\omega=\omega_{LS\pm}}\frac{T}{\epsilon_0(\omega - k_z v_d)} \frac{1}{\partial \epsilon_{zz}/\partial \omega \vert_\omega } \,.
\end{align}
The total (subluminal and supraluminal) spatial fluctuation spectrum is therefore given by the ballistic singularities
\begin{equation}\label{eq:int_ezz}
 \langle E_z E_z^*\rangle_{k_z} =\frac{T}{\epsilon_0} \frac{\omega_{p}^2\mu}{\omega_{p}^2 \mu + k_z^2c^2} \, ,
\end{equation}
which generalizes the nonrelativistic result of Ref. \cite[]{Langdon_1979}. 

As a test, we have numerically integrated Eq. \eqref{eq:intez2} over the subluminal region $\omega \in[-k_{c-}c,k_{c+}c]$ and compared the results
to the formula
\begin{align} \label{eq:int_sublumzz}
\langle E_z E_z^*\rangle ^{|\beta_\phi|<1}_{k_z} &= \frac{T}{\epsilon_0} \frac{\omega_{p}^2\mu}{\omega_{p}^2 \mu + k_z^2c^2} \nonumber \\
   &- \sum_{\omega=\omega_{LS\pm}}\frac{T}{\epsilon_0(\omega - k_z v_d)} \frac{1}{\partial \epsilon_{zz}/\partial \omega \vert_\omega } \, .
\end{align}
where the supraluminal eigenmodes $\omega_{LS}$ are obtained from solving Eq. \eqref{eq:supralumzz}. Figure \ref{fig:intez2_alpha0} shows
excellent agreement between the numerical and theoretical spectra for $\mu=(1,10,100)$ and $\beta_d=(0, 0.9)$. The discontinuities seen in the
subluminal spectra correspond to the supraluminal eigenmode cutoffs at $k=k_{c^\pm}$. Both negative and positive-phase velocity supraluminal
eigenmodes exist for  $k_z<k_{c-} <k_{c+}$, whereas only the positive-phase velocity supraluminal eigenmode exists for $k_{c-}<k_z<k_{c+}$ and all
eigenmodes are subluminal for $k_z>k_{c+}$. As a result, the total and subluminal spectra exactly coincide for $k_z>k_{c+}$. Note that these
discontinuities are hard to capture numerically in the low-temperature regime ($\mu=100$) due to the delta-like trace of the  subluminal eigenmode
for $|\beta_\phi| \sim 1$ [see Figs. \ref{ez2_spectrum_alpha0}(e,f)].

\subsection{Transverse fluctuations}

\subsubsection{Basic formulae}

Let us now consider the electromagnetic fluctuations propagating parallel to the plasma drift velocity. Plugging $\alpha=0$ into Eq. (\ref{eq:esilonxx}) first yields 
\begin{equation}\label{eq:epsxx}
  \epsilon_{xx}=1-\sum_{s}\frac{2\pi F_s \mu_{s}\omega_{ps}^2}{\omega^2} (\beta_{\phi}-\beta_{ds}) A_s(\beta_\phi).
\end{equation}
The electromagnetic spectra then write
\begin{align}\label{eq:fluctuxx}
\langle E_xE_x^*\rangle_{k_z,\omega} &= \frac{\langle j_xj_x^*\rangle_{k_z,\omega}}{\omega^2 \epsilon_0^2|\epsilon_{xx}-\frac{k_z^2c^2}{\omega^2}|^2} \\
  &= \frac{2}{\epsilon_0 |\epsilon_{xx}-\frac{k^2c^2}{\omega^2}|^2} \sum_s \frac{T_s \Im(\chi_{xx}^{(s)})}{\omega - k_z v_{ds}} \,, \\
\langle B_yB_y^*\rangle_{k_z,\omega} &=\frac{k_z^2}{\omega^2}  \langle E_x E_x^*\rangle _{k_z,\omega} \, .
\end{align}
This is the generalized fluctuation-dissipation theorem for a multispecies plasma with arbitrary drift velocities and temperatures. For equal
temperatures and drift velocities, it reduces to
\begin{align}
 \langle E_x E_x^*\rangle_{k_z,\omega} &= \frac{-2T}{\epsilon_0 (\omega - k_z v_{d})} \Im{\left(\frac{1}{\epsilon_{xx}-\frac{k_z^2c^2}{\omega^2}}\right)} \, , \label{eq:fluctuex2}\\
\langle B_y B_y^*\rangle_{k_z,\omega} &= \frac{-2T/v_\phi^{2}}{\epsilon_0(\omega - k_z v_{d})} \Im{\left(\frac{1}{\epsilon_{xx}-\frac{k_z^2c^2}{\omega^2}}\right)}\,. \label{eq:fluctuby2}
\end{align} 

\subsubsection{Dispersion relation}\label{part:exact_dispe_xx}

Combining Eqs. \eqref{eq:dispe2} and \eqref{eq:epsxx}, the dispersion relation of the transverse fluctuations can be recast as
\begin{align}\label{eq:supralumxx}
  k_z^2c^2 &= \frac{1}{\beta_{\phi}^2-1}\sum_s 2\pi F_s \mu_s \omega_{ps}^2 (\beta_{\phi}-\beta_{ds}) A_s(\beta_\phi) \nonumber \\
  & = \mathcal{H}(\beta_\phi)\, .
\end{align}
The transverse normal modes $\omega_T(k_z)$ can then be numerically computed using the method detailed in Sec. \ref{part:exact_dispe_zz}.
An analytical expression of the supraluminal transverse modes $\omega_{TS}$ can be obtained in the $k_z \to 0$ limit, which reads
\begin{equation}
  \omega_{TS}^2(0) =  \sum_s \omega_{ps}^2 \mu_s \langle \beta_x^2 \rangle_s \, ,
\end{equation}
where $\langle \beta_x^2 \rangle_s$ denotes the average of $\beta_x^2$ for species $s$. This limiting value is independent of the phase velocity
(see the eigenmode curves in Figs. \ref{ex2_spectrum_alpha0}(a-f)).

Taking $\lim_{\beta_\phi \to \pm 1}$ in Eq. \eqref{eq:supralumxx}, given $\lim_{\beta_\phi \to \pm 1} A(\beta_\phi)$ is finite and non-zero, yields $\lim_{\beta_\phi \to \pm 1} k_z^2 = \infty$ 
so that there are only exactly two transverse supraluminal modes (one per sign of $\beta_\phi$) as shown in Figs. \ref{ex2_spectrum_alpha0}(a-f).

\subsubsection{$(\omega,k)$-resolved spectrum}

\begin{figure*}
\centering
\includegraphics[scale=1]{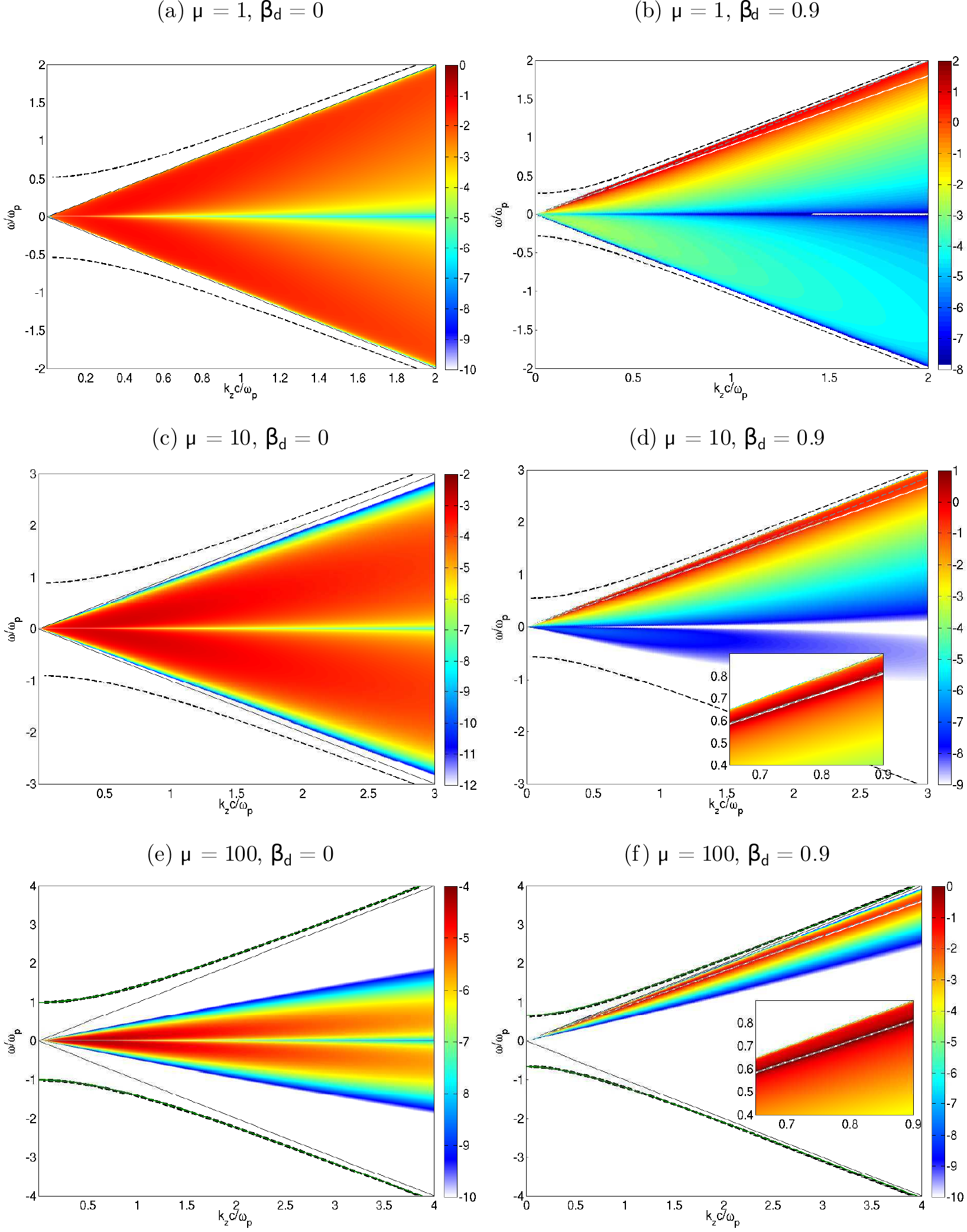}
\caption{\label{ex2_spectrum_alpha0}Power spectrum $\langle E_x E_x^*\rangle_{k_z,\omega}$ (normalized to $m_e^2c^3/e^2$) in $\log_{10}$ scale
for an electron/pair plasma and various $\mu$ and $\beta_d$ values. The two black solid lines correspond to $\beta_\phi = \pm 1$.
The dashed curves plot the exact eigenmodes computed from Eq. \eqref{eq:supralumxx}. In the low-temperature case ($\mu=100$),
the solid line plots the nonrelativistic tranverse mode in panel (e) and its Lorentz transformation in panel (f). The subpanels in
(d) and (f) show the exact eigenmode (grey dashed line) and the approximate beam mode $\omega=k_z\beta_d$ (white solid line).}
\end{figure*}

\begin{figure*}
\centering
\includegraphics[scale=1]{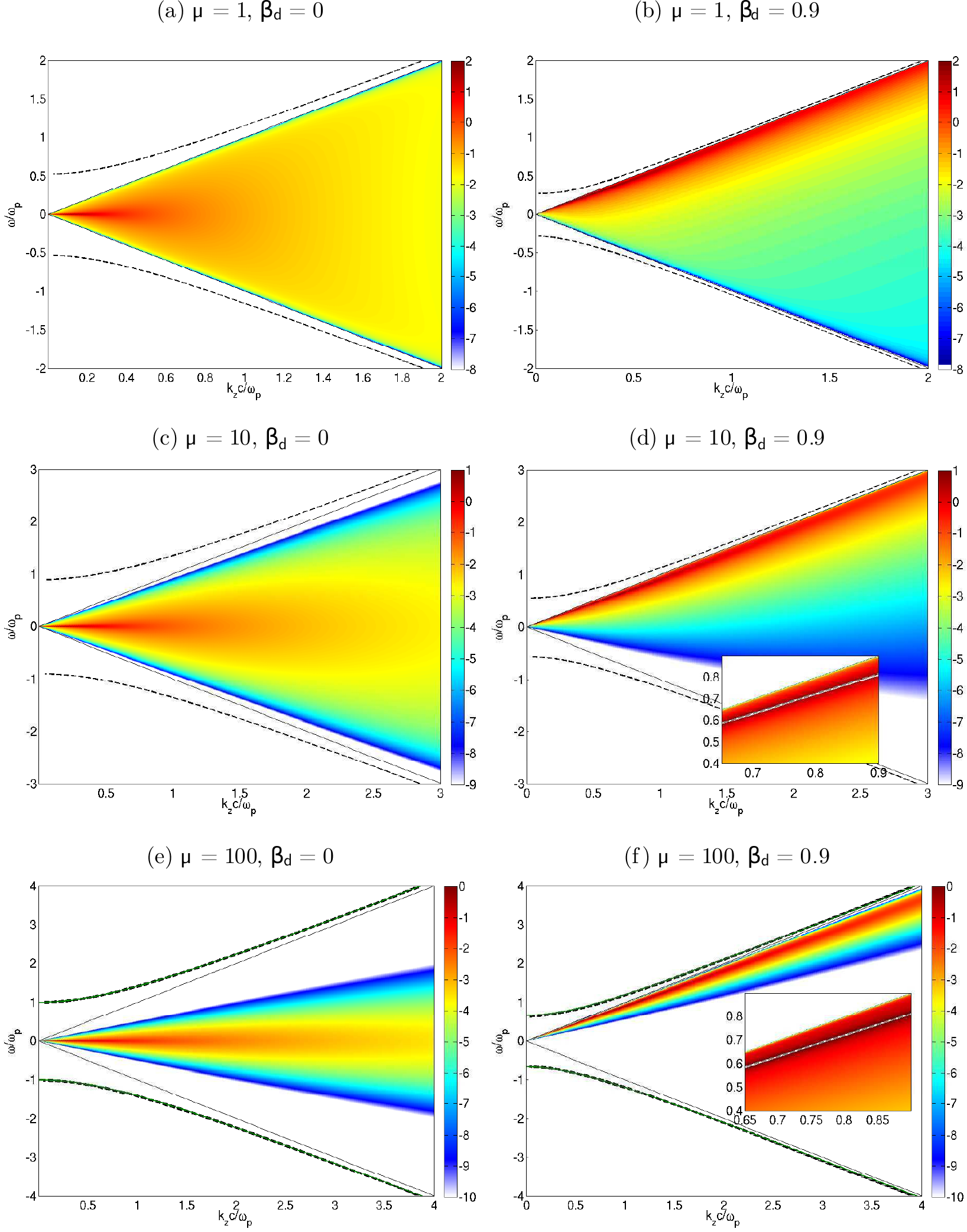}
\caption{\label{by2_spectrum_alpha0}Power spectrum $\langle B_y B_y^*\rangle_{k_z,\omega}$ (normalized to $m_e^2c/e^2$) in $\log_{10}$ scale
for an electron/pair plasma and various $\mu$ and $\beta_d$ values. The two black solid lines correspond to $ \beta_\phi  = \pm 1$.
The dashed curves plot the exact eigenmodes computed from Eq. \eqref{eq:supralumxx}. In the low-temperature case ($\mu=100$),
the solid line plots the nonrelativistic tranverse mode in panel (e) and its Lorentz transformation in panel (f). The subpanels in
(d) and (f) show the exact eigenmode (grey dashed) and the approximate beam mode $\omega=k_z\beta_d$ (white).}
\end{figure*}

\begin{figure}
\centering
\includegraphics[scale=1]{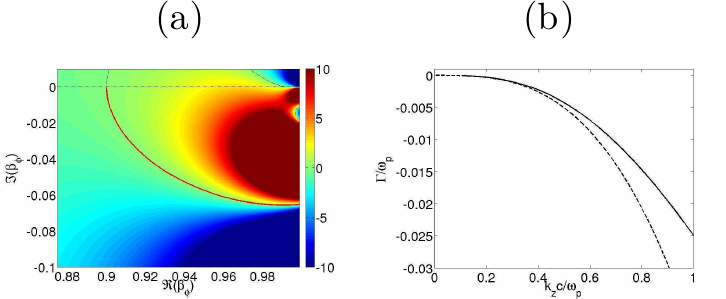}
\caption{\label{beam_mode} (a) Map of the complex function $\mathcal{H}(\beta_\phi)$ defined by Eq. \eqref{eq:supralumxx} for $\mu = 100$ and $\beta_d=0.9$:
the dashed line plots the isocontour $\Im \mathcal{H}=0$ and the solid line plots the $\Re \mathcal{H} > 0$ part of this isocontour, which corresponds
to the acoustic-like eigenmode shown in Fig. \ref{ex2_spectrum_alpha0}(f). (b) Damping rate of this eigenmode \emph{vs} $k_z$: comparison between the numerical
solution (solid line) and the approximate solution \eqref{eq2} (dashed curve).}
\end{figure}

Figures \ref{ex2_spectrum_alpha0}(a-f) and \ref{by2_spectrum_alpha0}(a-f) represent the spectra of the transverse electric and magnetic fluctuations, respectively,
for various values of $\mu$ and $\beta_d$. 
As for the longitudinal spectra, the supraluminal fluctuations are proportional to the delta function of the dispersion relation. Consequently, Eqs. \eqref{eq:fluctuex2}
and \eqref{eq:fluctuby2} vanish for $ \vert \beta_\phi \vert > 1 $ except along the supraluminal solutions of the dispersion relation Eq. \eqref{eq:supralumxx}. 
In the $\beta_d=0$ case, the subluminal spectra are symmetric 
with respect to $\omega=0$ and do no exhibit localized maxima due to
the absence of weakly-damped subluminal eigenmodes. By contrast, the subluminal spectra associated to $\beta_d=0.9$ are peaked close to a weakly-damped
acoustic-like branch $\omega_T \sim k_z v_d$. The damping rate of this so-called beam mode can be estimated by inserting $\omega_T = k_zv_d+i\Gamma(k)$
into Eq. \eqref{eq:dispe1} with $|\Gamma| \ll k_zv_d$. A first-order Taylor expansion of $\epsilon_{xx}(k_zv_d+i\Gamma,k)$ then yields 
\begin{equation}\label{eq:rtepsxx}
  \epsilon_{xx}(k_z\beta_d,k_z) +i\Gamma \frac{\partial \epsilon_{xx}}{\partial \omega}(k_z\beta_d,k_z) -\frac{k_z^2c^2}{(k_z v_d +i\Gamma)^2}=0.
\end{equation}
Taking the imaginary part of the above equation gives the damping rate
\begin{equation} \label{eq2}
  \Gamma(k_z)=-\beta_d^2\frac{k_z^3c^3}{\sum_s (2\pi)^2F_s\omega_{ps}^2 \mu_s f_A(\beta_d)}.
\end{equation}
Figure \ref{beam_mode}(b) shows that, for $\mu = 100$ and $\beta_d=0.9$, this expression closely matches the numerical solution up to $k_zc/\omega_p \sim 0.4$.

\subsubsection{$k$-resolved spectrum} \label{sec:trans_k_spec}

\begin{figure}
\centering
\includegraphics[scale=1]{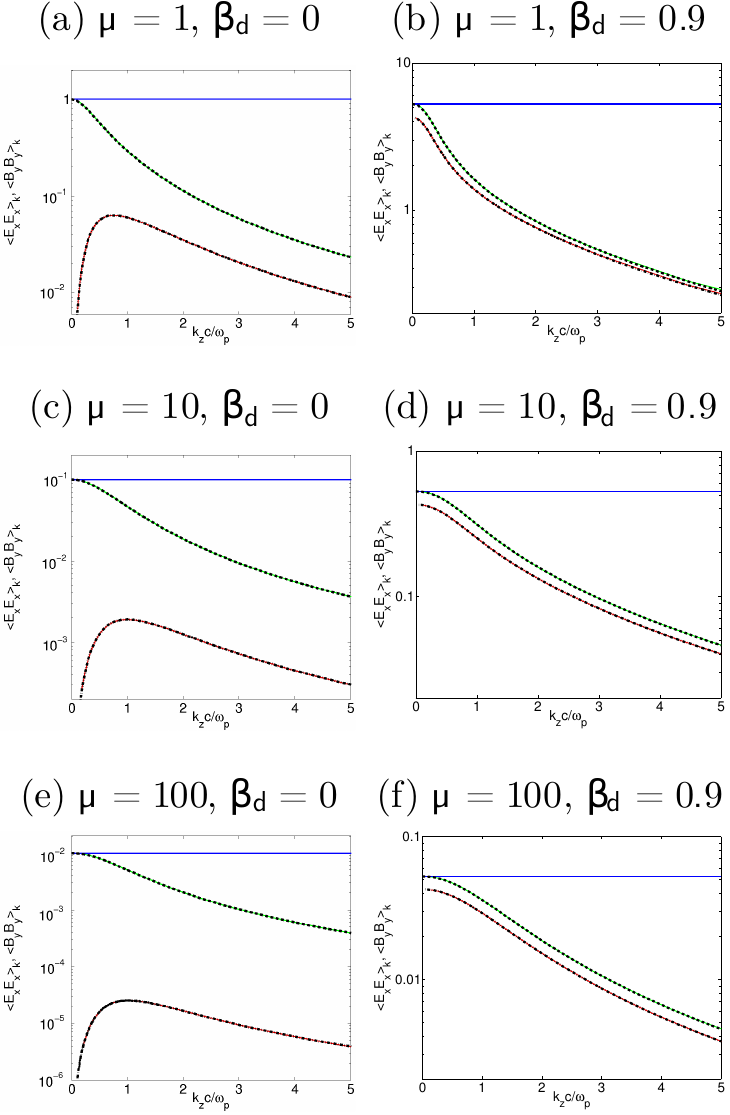}
\caption{\label{int_emg_alpha0} Spatial power spectra of the electromagnetic fluctuations  in an electron/pair plasma for for $\alpha=0$ and
various values of $\mu$ and $\beta_d$. The electric and magnetic spectra are normalized to $\frac{m_e^2c^3}{e^2}(\sum_s \omega_{ps}^2)^{1/2}$
and $\frac{m_e^3c^3 \epsilon_0}{n e^4}(\sum_s \omega_{ps}^2)^{1/2}$, respectively. Comparison of the total fluctuation spectra (blue solid line),
subluminal $\langle E_xE_x^*\rangle ^{|\beta_\phi|<1}_{k_z}$ spectra (red solid line), the subluminal $\langle B_y B_y^*\rangle ^{|\beta_\phi|<1}_{k_z}$
spectra (green solid line). Also plotted are the numerical integration of \eqref{eq:_int_ex2}  (dashed line) and of \eqref{eq:_int_by2}
(dashed-dotted line).}
\end{figure}

The $\omega$-integration of the transverse fluctuation spectra proceeds as in the electrostatic case. Let us therefore introduce the functions
\begin{align}
  I_E(k_z,\omega) &=  \frac{-2T}{\epsilon_0 (\omega - k_z v_{d})} \left(\frac{1}{\epsilon_{xx}-\frac{k_z^2c^2}{\omega^2}}-1\right)  \, ,  \label{eq:ie} \\
  I_B(k_z,\omega) &=  \frac{-2T}{\epsilon_0 (\omega - k_z v_{d})}\frac{1}{v_{\phi}^2 (\epsilon_{xx}-\frac{k_z^2c^2}{\omega^2})}  \, ,   \label{eq:ib} 
\end{align}
such that
\begin{align}
  \langle E_x E_x^*\rangle_{k_z} &=  \Im \int_{-\infty}^{+\infty} \frac{d\omega}{2\pi} I_E(\omega,k_z)\, , \label{eq:_int_ex2} \\
\langle B_y B_y^*\rangle_{k_z}  &= \Im \int_{-\infty}^{+\infty} \frac{d\omega}{2\pi} I_B(\omega,k_z) \, . \label{eq:_int_by2}
\end{align}
Making use of the closed contour $G$ (Fig. \ref{fig:contour_G}) and of the following limits
\begin{align}
  &\lim_{|\omega|\rightarrow 0}\frac{1}{\epsilon_{xx}-\frac{k_z^2c^2}{\omega^2}}-1 =0 \, , \\
  &\lim_{|\omega|\rightarrow 0}\frac{1}{\epsilon_{xx}-\frac{k_z^2c^2}{\omega^2}}\frac{1}{v_\phi^{2}} =0 \,,
\end{align}
we find that the supraluminal part of the transverse fluctuation spectrum is again given by the imaginary singularities related to the
transverse supraluminal eigenmodes $\omega_{TS}(k_z)$:
\begin{align}
 \langle E_x E_x^*\rangle ^{|\beta_{\phi}|>1}_{k_z}  &= -\sum_{\omega=\omega_{TS}}\frac{T}{\epsilon_0(\omega - k_z v_d )} \frac{1}{\frac{\partial
  \epsilon_{xx}}{\partial \omega}+2\frac{k_z^2c^2}{\omega^3}}   \label{eq:int_transv_supralume} \, , \\
 \langle B_y B_y^*\rangle ^{|\beta_{\phi}|>1}_{k_z}  &= -\frac{k_z^2}{\omega^2} \sum_{\omega=\omega_{TS}} \frac{T}{\epsilon_0 (\omega - k_z v_d )}  
  \frac{1}{\frac{\partial \epsilon_{xx}}{\partial \omega}+2\frac{k_z^2c^2}{\omega^3} }  \label{eq:int_transv_supralumb} \, .
\end{align}
Similarly, the total spatial fluctuations are determined by the ballistic singularities, and turn out to be independent of the wavenumber:
\begin{equation}\label{eq:int_transv}
\epsilon_0 \langle E_x E_x^*\rangle_{k_z}  = \frac{1}{\mu_0} \langle B_y B_y^*\rangle_{k_z}  = \gamma_d^2 T \, .
\end{equation}
In the case of a vanishing drift velocity, this formula is identical to the nonrelativistic expression \cite[]{Sitenko_fluctuations}, as pointed out
in Ref. \cite[]{Klimontovich_1982}.
The subluminal electric and magnetic spectra, $\langle E_x E_x^*\rangle^{\vert\beta_\phi\vert<1}_{k_z}$ and $\langle B_y B_y^*\rangle^{\vert\beta_\phi\vert<1}_{k_z}$,
readily follow from subtracting Eq. \eqref{eq:int_transv_supralume} and \eqref{eq:int_transv_supralumb} to Eq. \eqref{eq:int_transv}. Simple analytical expressions
can be obtained at $k_z=0$:
\begin{align}
 \langle E_x E_x^*\rangle^{\vert \beta_\phi \vert<1}_{k_z=0}  &= \frac{\gamma_d^2 \beta_d^2}{\epsilon_0} T \, , \\
\langle B_y B_y^*\rangle^{\vert \beta_\phi \vert <1}_{k_z=0} &= \mu_0 \gamma_d^2 T\, . \label{eq:ByBy_0_trans} 
\end{align}
Whereas the $k_z=0$ magnetic spectra is always purely subluminal, its electric counterpart involves, in general, both supraluminal and subluminal
contributions. The latter vanishes for $\beta_d$ = 0 and prevails in the ultra-relativistic limit $\gamma_d \gg 1$. The full $k_z$-dependence of the
subluminal, supramuminal and total spectra of the transverse fluctuations is shown in Figs. \ref{int_emg_alpha0}(a-f) 
for various values of  $\mu$ and $\beta_d$.

\section{Magnetic fluctuations with wave vectors normal to the plasma drift velocity}\label{sec:mag_fluc}

\subsection{Basic formulae}

We now consider the spectrum of magnetic fluctuations propagating normally to the mean plasma velocity ($\alpha=\pi/2$). Without loss
of generality, the wave vector is taken along the $y$-axis. Since we are interested in estimating the seed of growing filamentation modes
in counterpropagating plasma flows \cite[]{Bret_2013}, the magnetic field is chosen along the $x$-axis, so that the electric field lies in
the $yz$-plane. In contrast to the previous cases, the tensorial quantities $\boldsymbol{\epsilon}$, $\mathbf{Z}$, $\langle\mathbf{jj}^\dagger\rangle$
and $\langle \mathbf{BB}^\dagger\rangle$ are now no longer diagonal. Their expressions for a Maxwell-J\"uttner-distributed, multispecies plasma
are given in Appendix \ref{ap:alpha_pis2}. For an electron or pair plasma with equal temperatures and velocities, one gets  
\begin{align}
\langle j_xj_x^{*}\rangle_{k_y,\omega}&=\frac{2T\omega \Im{(\epsilon_{xx})}}{\epsilon_0} \,, \label{eq:sourcepis2xx}\\
\langle j_yj_y^{*}\rangle_{k_y,\omega}&=\frac{2T\omega \Im{(\epsilon_{yy})}}{\epsilon_0} \,, \label{eq:sourcepis2yy}\\
\langle j_zj_z^{*}\rangle_{k_y,\omega}&=\frac{2T\omega \Im{(\epsilon_{zz})}}{\epsilon_0} \nonumber \\
&= \frac{2T\Im{(\omega^2 \epsilon_{zz}- k^2c^2)}}{\epsilon_0\omega}\label{eq:sourcepis2zz} \, , \\
\langle j_yj_z^{*}\rangle_{k_y,\omega}&=-\frac{2T\omega \Im{(\epsilon_{yz})}}{\epsilon_0} \label{eq:sourcepis2yz} \,.
\end{align}
Inserting Eq. \eqref{eq:sourcepis2yy}-\eqref{eq:sourcepis2yz} into Eq. \eqref{eq:fluctubx05pi0} gives the magnetic spectrum
\begin{equation}\label{eq:fluctubx05pi}
 \langle B_xB_x^*\rangle _{k_y,\omega} = -\frac{2k_y^2T}{\epsilon_0\omega}
 \Im \left[\frac{\epsilon_{yy}}{(\omega^2 \epsilon_{zz}-k_y^2c^2)\epsilon_{yy}-\omega^2 \epsilon_{yz}^2 } \right ] \,.
\end{equation} 

\subsection{Dispersion relation}

The denominator of Eq. \eqref{eq:fluctubx05pi} corresponds to the electromagnetic dispersion \eqref{eq:dispe1} with $k_z=0$. Its explicit form is
obtained by plugging Eqs. \eqref{eq:epsilonp05ixx} and \eqref{eq:epsilonp05iyz}. There follows a second-order polynomial equation in $k_y^2$:
\begin{widetext} 
\begin{align}
  k_y^4c^4(\beta_{\phi}^2-1)-
 k_y^2c^2\left[(\beta_{\phi}^2-1)\sum_s 2\pi F_s \mu_s \omega_{ps}^2 \frac{B_s(\beta_\phi)}{v_{\phi}}- 
  \sum_s \mu_s \omega_{ps}^2 \beta_{ds}^2 +\sum_s 2\pi F_s\mu_s\omega_{ps}^2v_{\phi}A_s(\beta_\phi)\right]\nonumber\\
  +\sum_s 2\pi F_s \mu_s \omega_{ps}^2 \frac{B_s(\beta_\phi)}{v_{\phi}} \left[\sum_s 2\pi F_s\mu_s\omega_{ps}^2v_{\phi}A_s (\beta_\phi)-\sum_s \mu_s \omega_{ps}^2 \beta_{ds}^2 \right]
  -\left[\sum_s 2\pi F_s\mu_s\omega_{ps}^2C_s(\beta_\phi)\right]^2=0 \,. \label{eq:alphapis2dispe}
\end{align}
\end{widetext}
The solutions of the above equation then pertain to two distinct branches, $\beta_{\phi,1}=\omega_1/k_yc$ and $\beta_{\phi,2}=\omega_2/k_yc$,
defined, respectively, by
\begin{align}
  k_y^2 c^2 &=\frac{-a_1(\beta_{\phi,1}) + \sqrt{\Delta(\beta_{\phi,1}})}{2(\beta_{\phi,1}^{2}-1)}\label{eq:sol1} \, ,\\
  k_y^2 c^2 &=\frac{-a_1(\beta_{\phi,2}) - \sqrt{\Delta(\beta_{\phi,2)}}}{2(\beta_{\phi,2}^{2}-1)}\label{eq:sol2} \, .
\end{align}
We have introduced $\Delta=a_1^2-4(v_\phi^2-1)a_0$, where $a_n$ is the $n$th order coefficient of the polynomial in $k_y^2$ defined by
Eq. \eqref{eq:alphapis2dispe}. Again, we employ the Fried and Gould scheme to compute the entire set of eigenmodes associated to each branch.

We find that there exist a maximum number of four undamped, or weakly-damped, eigenmodes (two symmetric modes of opposite phase velocities per branch).
As displayed by the coloured dashed curves in  Figs. \ref{bx2_spectrum_alphapis2}(a-c), these solutions correspond to the Lorentz-transformed
dominant electromagnetic (branch 1) and electrostatic (branch 2) modes in the plasma rest frame. This is demonstrated in the low-temperature regime
($\mu=100$) by the precise coincidence between the exact curves and the Lorentz transforms of the modes $\omega=\omega_p(1+k^2c^2/\omega_p^2)^{1/2}$ and
$\omega=\omega_p(1+3k^2c^2/\omega_p^2\mu)^{1/2}$. Note that the eigenmodes associated to branch 1 are purely supraluminal for all $k_y$'s. 

Owing to the non-vanishing $\epsilon_{yz}$ term in Eq. \eqref{eq:dispe1}, the electric field associated to eigenmodes with wave vectors normal to the plasma
drift velocity has both longitudinal ($E_y$) and transverse ($E_z$) components. This property has been analyzed in detail in Refs. \cite[]{Tzoufras_2006,
Bret_2007a} in the context of the filamentation instability. The orientation of the electric field is determined by the following formula \cite[]{Bret_2004}:
\begin{equation}\label{eq:angleyz}
  \frac{E_y}{E_z}=-\frac{\omega^2\epsilon_{zz}-k_y^2c^2}{\omega^2\epsilon_{yz}}=-\frac{\omega^2 \epsilon_{yz}}{\omega^2\epsilon_{yy}} \, .
\end{equation}
Making use of Eqs. \eqref{eq:epsilonp05ixx}-\eqref{eq:epsilonp05iyz}, this can be recast as
\begin{align}\label{eq:angleyz2}
\frac{E_y}{E_z} &=
- \frac{\sum_s\pi F_s\omega_{ps}^2\mu_s \beta_\phi C(\beta_\phi)}{\omega^2-\sum_s2\pi F_s\omega_{ps}^2\mu_s \beta_\phi B(\beta_\phi)}
 \nonumber\\
&= \frac{k^2c^2-\omega^2+\sum_s\omega_{ps}^2\mu_s [2\pi F_s\beta_\phi A(\beta_\phi)-\beta_d^2]}{\sum_s2\pi F_s\omega_{ps}^2\mu_s \beta_\phi C(\beta_\phi)}
\end{align}
The above equation can be analytically evaluated in the $k_y \rightarrow0$ and $k_y \rightarrow\infty$ limits. It is easy to demonstrate that
\begin{align}
  \lim_{k_y \rightarrow \infty} \omega^2_{1\pm} &=\lim_{k_y \rightarrow \infty} \omega^2_{2\pm}=k_y^2c^2  \,, \label{eq:lim1}\\
  \lim_{k_y \rightarrow 0} \omega^2_{1\pm} &=\sum_s 2\pi F_s\omega_{ps}^2 \mu_s\int_{-1}^{+1} f_B \,d\beta \, , \label{eq:lim2}\\
  \lim_{k_y \rightarrow 0} \omega^2_{2\pm} &=\sum_s 2\pi F_s\omega_{ps}^2 \mu_s\int_{-1}^{+1}f_A \,d\beta  \nonumber\\
  &-\sum_s\omega_{ps}^2 \mu_s \beta_{ds}^2 \,.\label{eq:lim3}
\end{align}
where use has been made of $\lim_{\beta_\phi \rightarrow \infty}\beta_\phi B(\beta_\phi)= \int_{-1}^{+1} d\beta f_B$.
It follows that the eigenmode $\omega_{1,\pm}$ is purely longitudinal for $k_y=0$ ($E=E_y$) and becomes purely transverse for $k_y \rightarrow \infty$
($E=E_z$). By contrast, the eigenmode $\omega_{2,\pm}$  is purely longitudinal for $k_y \rightarrow \infty$ and purely transverse for $k_y=0$.
The $k_y$-dependence of the angle $\phi=\arctan(E_y/E_z)$ is plotted for both branches in Fig. \ref{int_bx2_alphapis2}(d) in a relativistically hot
case ($\mu=1$, $\beta_d=0.9$). 

\subsection{$(\omega,k)$-resolved spectrum}

\begin{figure}
\centering
\includegraphics[scale=1]{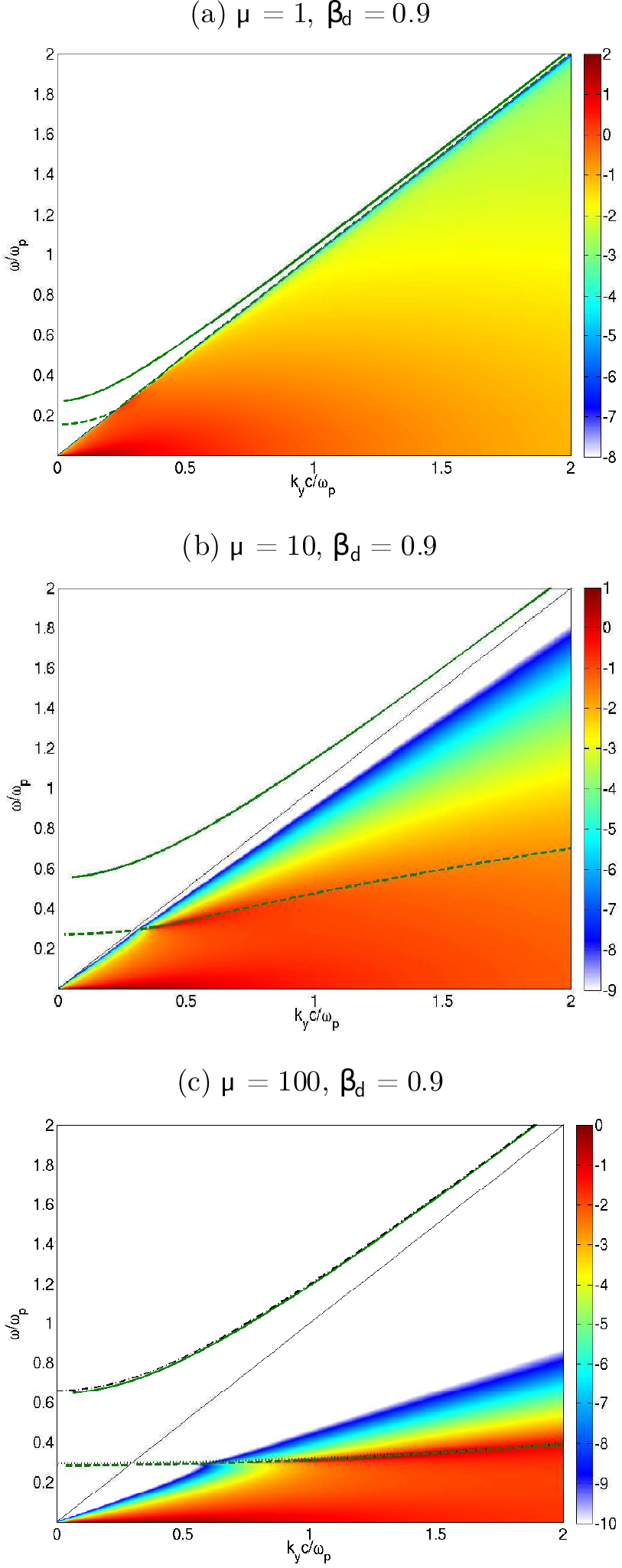}
\caption{\label{bx2_spectrum_alphapis2} Power spectrum $\langle B_xB_x^*\rangle _{k_y,\omega}$ (normalized to $m_e^2c/e^2$ and in $\log_{10}$
scale) of an electron or pair plasma for $\alpha=\pi/2$, $\beta_d=0.9$ and various values of $\mu$. The white solid lines delimit the subluminal
region $ \vert \beta_\phi \vert \le 1$. The exact eigenmodes are plotted in green solid lines (branch 1) and dashed curves (branch 2). In
panel (c) are also plotted the Lorentz transforms of the classical transverse eigenmode (black blue dotted-dashed line) and of the longitudinal
eigenmode (dotted black kine).}
\end{figure}

Figures \ref{bx2_spectrum_alphapis2}(a,b,c) display Eq. \eqref{eq:fluctubx05pi} for $\beta_d=0.9$ and $\mu=(1,10,100)$. Only half the spectrum
is shown due to its parity in $\omega$. As expected, the trace of the eigenmode $\omega_{2,+}$ increasingly stands out in the subluminal region as
the plasma temperature drops.
 
Similarly to Figs. \ref{by2_spectrum_alpha0}(a,c,e), these spectra exhibit a peaked structure around $\omega=0$. As pointed out in Refs.
\cite[]{Yoon_2007, Tautz_2007, Bret_2013}, these zero-frequency fluctuations are associated to the unstable magnetic filamentation (Weibel)
mode. Combining Eqs. \eqref{eq:epsilonp05ixx}-\eqref{eq:fluctubx05pi0} gives the closed-form expression
\begin{equation}
\langle B_xB_x^* \rangle_{k_y,\omega=0}=\frac{\gamma_d^3}{k^3}\frac{\mathcal{J}}{\left(k^2 + \frac{\omega_p^2 \mu}{\gamma_d^2}\right)^2}
\frac{\omega_p^2e^{-\frac{\mu}{\gamma_d}}}{\mu^2 K_2(\frac{\mu}{\gamma_d})} \, ,
\end{equation}
with
\begin{align}
 \mathcal{J}=&(k_y^2+\omega_p^2\mu)^2 \left[\left(\frac{\mu}{\gamma_d}+1 \right) \left(1+2\beta_d^2 \right) + \beta_d^2\frac{\mu^2}{\gamma_d^2} \right]  \nonumber\\
 &+\omega_p^4\mu^2\beta_d^2 \left[\left(\frac{\mu}{\gamma_d}+1 \right) \left(2+\beta_d^2 \right) + \frac{\mu^2}{\gamma_d^2} \right]  \nonumber\\
 &-2\left(k_y^2+\omega_p^2\mu \right)\omega_p^2\mu\beta_d^2 \left( 3 \frac{\mu}{\gamma_d}+3 +\frac{\mu^2}{\gamma_d^2} \right) \, .
\end{align}
We have thus generalized to arbitrary temperatures and drift velocities the formula obtained by Yoon \cite{Yoon_2007} for a nondrifting,
nonrelativistic Maxwellian plasma. As in the nonrelativistic limit, we find that $\langle B_xB_x^*\rangle_{k_y,\omega=0}$ scales as $k^{-3}$
for $k_y^2 \ll \omega_p^2\mu/\gamma_d^2$. 

\subsection{$k$-resolved spectrum} \label{sec:mag_k_spec}

In order to carry out the integration of $\langle B_x B_x^* \rangle_ {k_y,\omega}$ over $\omega \in \mathbb{R}$, we define the function
\begin{equation}\label{eq:lb}
  L_B =-\frac{2T}{\epsilon_0\omega}
  \frac{k^2 \epsilon_{yy}}{(\omega^2 \epsilon_{zz}-k_y^2c^2)\epsilon_{yy}-\omega^2 \epsilon_{yz}^2 } \, ,
\end{equation}
so that 
 \begin{equation}\label{eq:int_alphas2_bx2} 
\langle B_x B_x^* \rangle_{k_y} = \Im \int_{-\infty}^{+\infty} \frac{d\omega}{2\pi} L_B \, .
 \end{equation}
Proceeding as in Secs. \ref{sec:long_k_spec} and \ref{sec:trans_k_spec}, we find that the total $k$-resolved magnetic spectrum is determined by 
the $\omega^{-1}$ term of Eq. \eqref{eq:lb}. The calculation of the corresponding residue requires to evaluate the $\omega\rightarrow 0$ limits
of $\epsilon_{yy}$, $\omega \epsilon_{yz}$ and $\omega^2 \epsilon_{zz}$. These calculations are performed in Appendix \ref{sec:w0_mag_spectrum}.
Using Eqs. \eqref{eqc:2yy}-\eqref{eqc:2zz}, the total spectrum reads
\begin{equation}\label{eq:int_bx2_pis2}
\langle B_x B_x^* \rangle_{k_y} =\mu_0T\frac{k_y^2c^2+\omega_p^2 \mu}{k_y^2c^2 + \omega_p^2 \mu/\gamma_d^2}\, .
\end{equation}
As usual, the supraluminal spectrum results from the supraluminal singularities $\omega_{1,2S}$ of Eq. \eqref{eq:lb}. 
\begin{equation}\label{eq:int_bx2_pis2_suparlum}
 \langle B_x B_x^*\rangle ^{|\beta_\phi|>1}_{k_y} = \sum_{\omega = \omega_{1,2S}}\mu_0 T
  \frac{k^2 \epsilon_{yy}}{\omega \partial D/\partial \omega \vert_\omega}\, ,
\end{equation}
with $D=(\omega^2 \epsilon_{zz}-k_y^2c^2)\epsilon_{yy}-\omega^2 \epsilon_{yz}^2 $. For $k_y^2c^2 > k_{2S}^2$ (Eq. \eqref{eq:k2S}), only the pair
of solutions $\omega_{1S\pm}$ contribute to the above equation, yielding $\langle B_x B_x^*\rangle ^{|\beta_\phi|>1}_{k_y} \sim \mu_0 T$.
For the sake of completeness, the $\langle \mathbf{E} \mathbf{E}^\dagger\rangle_{k_y}$ and $\langle B_z B_z^*\rangle_{k_y}$ spectra are
summarized in Appendix \ref{sec:ExEx_EyEy_EzEz_BzBz}.

Figures \ref{int_bx2_alphapis2}(a,b,c) plot the $k_y$-dependence of the total, supraluminal and subluminal magnetic spectra for
$\beta_d=0.9$ and various values of $\mu$. Note that the magnetic fluctuations are purely subluminal at $k_y=0$, with
$\langle B_x B_x^*\rangle_{k_y=0}=\mu_0\gamma_d^2 T$.  As expected in the infinite-wavelength limit, we retrieve the value
\eqref{eq:ByBy_0_trans} obtained for $\alpha = 0$. Again, the numerical integration of Eq. \eqref{eq:fluctubx05pi} over the
subluminal $\omega$-domain accurately reproduces the analytical formula deduced from subtracting Eq. \eqref{eq:int_bx2_pis2_suparlum}
to Eq. \eqref{eq:int_bx2_pis2}. The jumps seen in the supraluminal and subluminal spectra stem from the pair of solutions
$\omega_{2S\pm}$ turning subluminal above a critical wave vector 
\begin{equation}\label{eq:k2S}
k_{2S}^2 c^2= \sum_s \frac{\omega_{ps}^2}{\gamma_d^3}
  \frac{K_1(\frac{\mu_s}{\gamma_{ds}}) + 2\frac{\gamma_{ds}}{\mu_s}K_0(\frac{\mu_s}{\gamma_{ds}})}{K_2(\frac{\mu_s}{\gamma_{ds}})} \, ,
\end{equation}
obtained from a Lorentz transform of Eq. \eqref{eq:kc} taken at $\beta_d=0$. As in Sec. \ref{sec:long_k_spec}, this
discontinuity is difficult to resolve in the cold case $\mu = 100$.

\begin{figure}
\centering
\includegraphics[scale=1]{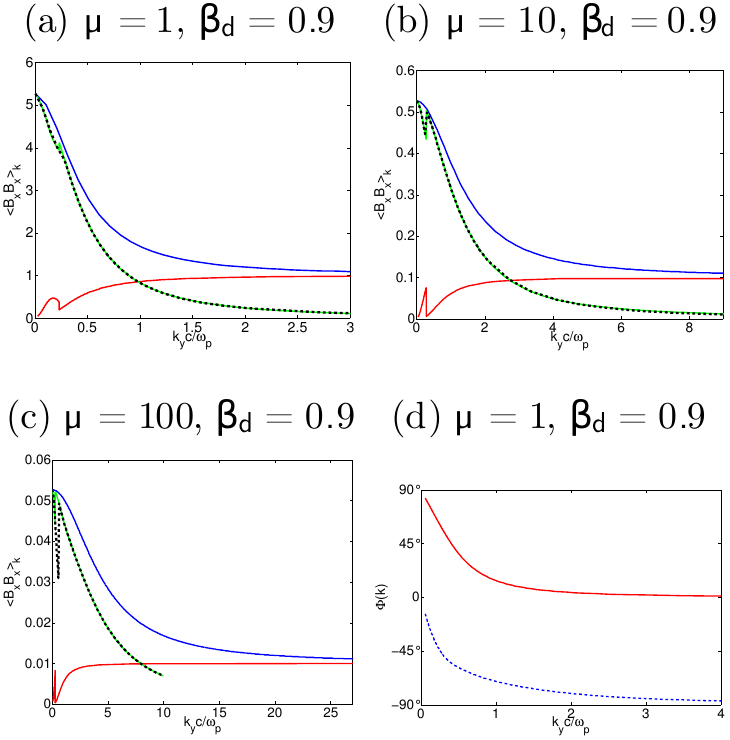}
\caption{ \label{int_bx2_alphapis2} (a,b,c) $<B_x B_x^*>_{k_y}$ spectra (in $\frac{m_e^2c^3}{e^2}(\sum_s \omega_{ps}^2)^{1/2}$ units)
of a single-temperature electron/pair plasma for $\beta_d = 0.9$ and various values of $\mu$. Comparison of the total (blue), supraluminal (red) and
subluminal (green) spectra. The dashed black curve plots the numerical integration of Eq. \eqref{eq:fluctubx05pi} over the subluminal domain
$\vert \omega \vert < k_yc$. (d) Variation of the angle $\phi(k_y)=\arctan(E_y/E_z)$ for the eigenmodes belonging to branch 1 (solid red) and branch 2 (dashed blue).}
\end{figure}

\section{Comparison to PIC simulations} \label{sec:PIC_sims}

In the following, we confront our analytical formulae to the electromagnetic fluctuations induced in a numerical PIC-modeled plasma. The purpose
is to analyze the numerical noise seeding amplified modes in PIC simulations of relativistic plasma instabilities. Our code \cite[]{Lefebvre_2003}
employs the standard Yee solver for the Maxwell equations \cite[]{Yee_1966} and a charge-conserving current deposition scheme \cite[]{Esirkepov_2001}
with a third-order weight factor for the macro-particles. The simulation geometry is one-dimensional in space (along or normal to the plasma velocity)
and three-dimensional in momentum. The spatial and temporal step sizes are $\Delta z = 0.1c/\omega_p$ and $\Delta t = 0.095/\omega_p$, respectively.
The plasma length is $560c/\omega_p$ with periodic boundary conditions for both the fields and macro-particles. The fields are initially zero. We
consider an $e^+e^-$ pair plasma initialized according to Eq. \eqref{eq:mj}, with $\mu=10$ and $\beta_d = 0.9$. The macro-particles have a charge
and a mass equal, respectively, to $Q_p = \pm W_p q_e$ and $M_p = W_p m_e$, where $W_p$ is the statistical weight. For the numerical plasma
to behave collectively as its physical counterpart, the plasma frequencies of the two systems must be equal, which implies
\begin{equation}
  W_p = \frac{m_e \epsilon_0 \omega_{pe}^2}{e^2} \frac{\Delta z}{N_p} \,,
\end{equation}
where $N_p$ is the number of macro-particles per mesh and species ($N_p=2000$ here). In the 1-D geometry under consideration, $W_p$ thus
corresponds to an areal density. Since the normalized inverse temperature $\mu$ is an invariant, the susceptibility tensor is unchanged
in the simulation. By contrast, the source term is modified according to
\begin{equation}\label{sroucepic}
 \langle \mathbf{jj}^\dagger\rangle_{\mathbf{k},\omega}^\mathrm{PIC}= W_p^{-1} \langle \mathbf{jj}^\dagger \rangle_{\mathbf{k},\omega} \,.
\end{equation}
The effect of the finite spatial width of the macro-particles is here neglected \cite[]{Birdsall_1985}.
The relevant simulated quantity to be compared to the theoretical spectrum is therefore $W_p \langle \mathbf{EE}^\dagger \rangle_{\mathbf{k},\omega}^\mathrm{PIC}$.
In practice, the fluctuation power spectrum is computed from the absolute square of the fast Fourier transform in space and time $\vert \mathrm{FFT}_{z,t}(E_\alpha)\vert^2$.
Care is taken to select a temporal domain over which the system has reached a quasi-stationary state. 
The total simulation time is a few $1000 \omega_{pe}^{-1}$

\begin{figure}[!]
\centering
\includegraphics[scale=1]{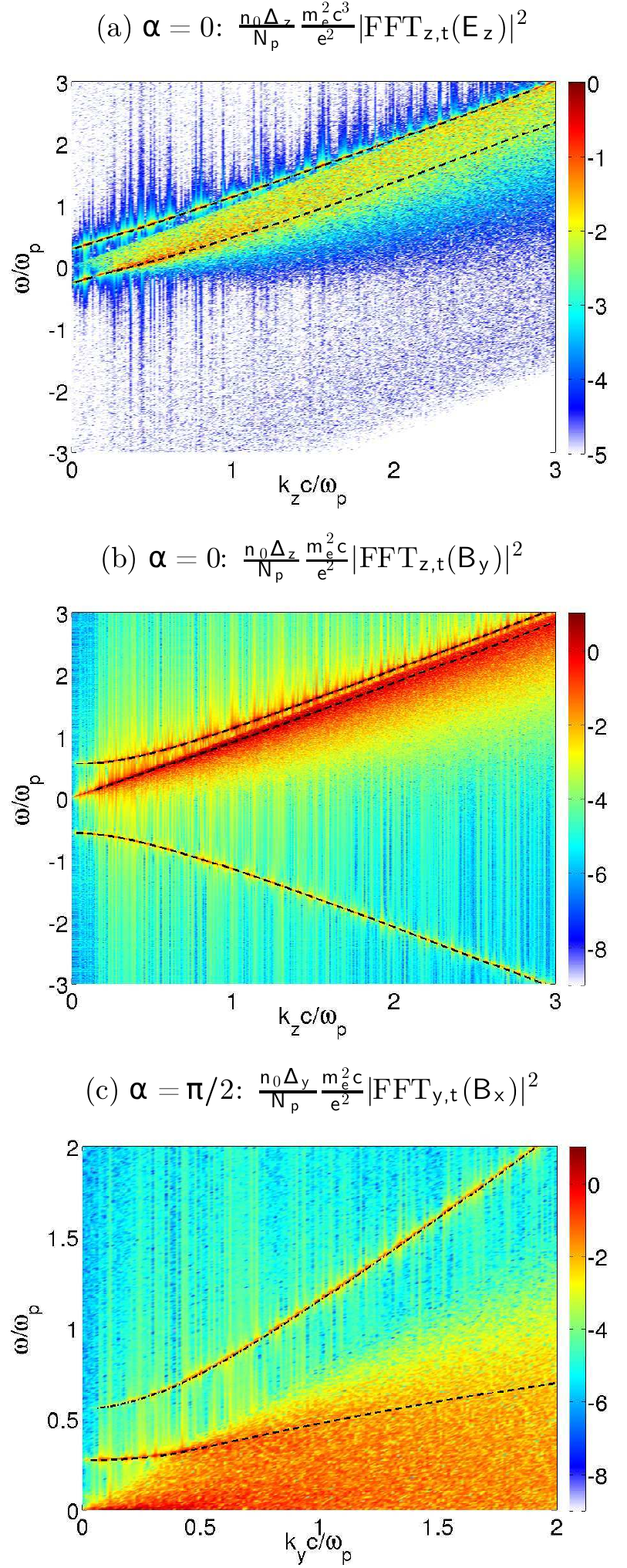}
\caption{\label{fig:PIC_fluctu} Simulated $(k,\omega)$ spectra in $\mathrm{log}_{10}$ scale for $\mu=10$ and $\beta_d =0.9$ (see text for details).
Panels (a) and (b) correspond to wave vectors parallel to the beam ($\alpha =0$), whereas panel (c) corresponds to wave vectors normal to the beam
($\alpha = \pi/2$). The dashed and dotted-dashed curves plot the eigenmodes solving Eq. \eqref{eq:supralumzz} (a), Eq. \eqref{eq:supralumxx} (b) and
Eq. \eqref{eq:alphapis2dispe} (c). The system size is $L=560c/\omega_p$ in all cases.}
\end{figure}

Figures \ref{fig:PIC_fluctu}(a,b,c) display the $(k,\omega)$-resolved power spectra of the simulated electric and magnetic fluctuations for $\alpha =0$ (a,b)
and $\alpha=\pi/2$ (c). In the latter case, the simulation resolves the beam-normal $y$-axis. These results satisfactorily agree (over $\sim 4$ decades)
with the theoretical predictions depicted in Figs. \ref{ez2_spectrum_alpha0}(d), \ref{by2_spectrum_alpha0}(d) and \ref{bx2_spectrum_alphapis2}(b). 
A noticeable difference is that the supraluminal eigenmodes appear as finite-width structures in the simulations instead of delta-like singularities.
This can be attributed to a number of reasons: the finite temporal window, the numerical collisions between the finite-width macro-particles and the non-adiabatic
switch-on of the fields. Regarding the latter, the level of the supraluminal fluctuations was theoretically shown to be highly sensitive to the details
of the plasma initialization in Refs. \cite[]{Lerche_1968,Lerche_1969b,Lerche_1969c}. A PIC study of the impact of the plasma initialization upon the
asymptotic field fluctuations is outside the scope of this paper.

\begin{figure*}
\centering
\includegraphics[scale=1]{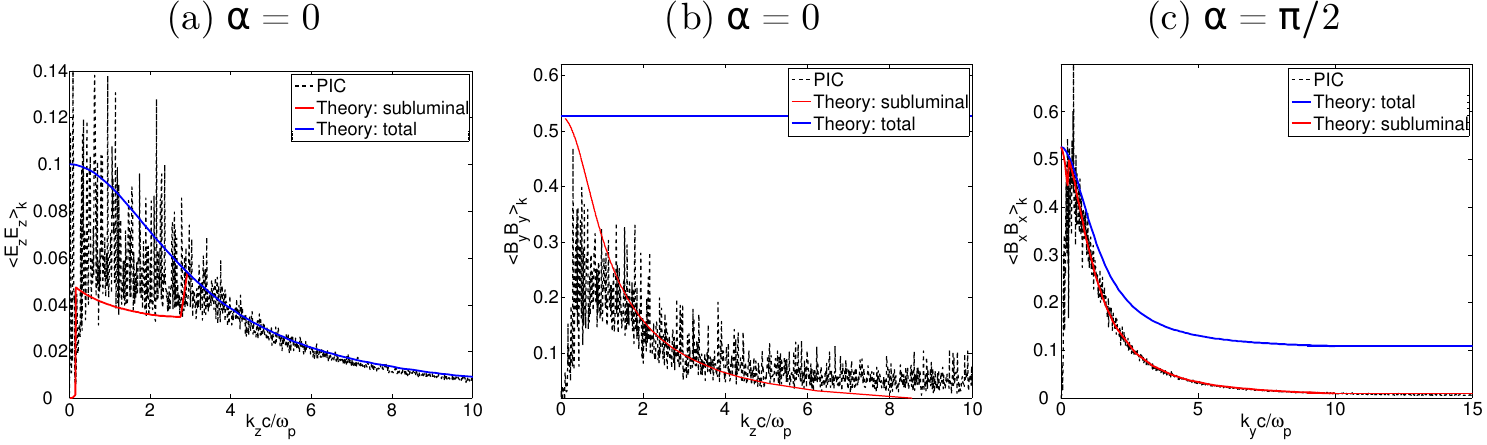}
\caption{\label{fig:int_PIC_fluctu} Spatial fluctuation spectra of $\langle E_zE_z^*\rangle_{k_z}$ for $\alpha =0$ (a), of $\langle B_yB_y^*\rangle_{k_z}$
for $\alpha =0$ (b) and $\langle B_xB_x^*\rangle_{k_y}$ for $\alpha =\pi/2$ (c): comparison of PIC simulation results (dashed line) and
theoretical subluminal (red solid lines) and total (blue solid lines) spectra (in $\frac{m_e^2c^3}{e^2} (\sum_s \omega_{ps}^2)^{1/2}$ units).
}
\end{figure*}

Let us now consider the spatial fluctuation spectra. From the same reasoning as above, one has to compare the theoretical formulae (obtained in
Secs. \ref{sec:long_k_spec}, \ref{sec:trans_k_spec} and \ref{sec:mag_k_spec}) to the simulated quantity $W_p\langle\vert\mathrm{FFT}_z(E_\alpha)\vert ^2\rangle_t$,
where $\langle \rangle_t$ denotes a temporal average. The PIC spatial spectra shown in Figs. \ref{fig:int_PIC_fluctu}(a,b,c) turn out to better match
the subluminal spectra than the total spectra. This behavior, which is particularly pronounced for the transverse fluctuations with $\alpha =0$
[Fig. \ref{fig:int_PIC_fluctu}(b)], confirms the observed discrepancy between the simulated and theoretical supraluminal spectra. We have checked
that the energy stored in the supraluminal structures of Figs. \ref{fig:PIC_fluctu}(a,b,c) significantly underestimates the theoretical expectations.

\section{Summary and conclusions} \label{sec:conclusions}

The power spectra of the electric and magnetic fluctuations spontaneously induced in unmagnetized, collisionless plasmas described by relativistic
Maxwell-J\"uttner distribution functions have been explicitly evaluated for wave vectors parallel or normal to the plasma mean velocity. Closed-form
analytical formulae of the $\omega$-integrated spectra have been worked out in all cases, distinguishing between the contributions of the subluminal
and supraluminal electromagnetic fluctuations. We have found that the well-known nonrelativistic results \cite[]{Akhiezer_fluctuations} still hold
for nondrifting relativistic plasmas. In passing, we have generalized the Fried and Gould method \cite[]{Fried_Gould_1961} to solve for the full
set of eigenmodes of the system. This technique, evidently, could also handle unstable systems and oblique wave vectors.

The particular case of magnetic fluctuations with wave vectors normal to the beam velocity has been treated in detail. We have found that the
long-wavelength ($k_y \ll \omega_p\sqrt{\mu}/c$) spatial magnetic fluctuations exceed the short-wavelength fluctuations by a factor $\gamma_d^2$.
An explicit expression of the $\omega =0$ magnetic fluctuations is also provided. This formula has recently served to estimate the seed and saturation
time of the relativistic filamentation instability of counterpropagating plasmas. Fair agreement with PIC simulations has been found provided both
plasmas are in thermal equilibrium before overlapping \cite[]{Bret_2013}.

Finally, we have confronted our theoretical results to 1-D PIC simulations of drifting thermal plasmas. Overall, the theoretical spectra are well reproduced in the
subluminal region. The eigenmode dispersion relations are accurately captured, yet with somewhat underestimated energy in the supraluminal domain.

\section*{Acknowledgments}
The authors gratefully acknowledge Antoine Bret for interesting discussions.
The PIC simulations were performed using HPC resources at TGCC/CCRT (Grant No. 2013-052707).

\appendix

\section{Derivation of Eq. \ref{eq:epszz}}
%\section{Derivation of Eq. $(29)$}
\label{ap:alpha_0} 

For $\alpha=0$, the $zz$ component of Eq. \eqref{eq:tenseur_eps} can then be recast as
\begin{align}
  \epsilon_{zz}=1+\sum_s \frac{\omega_{ps}^2 }{\omega^2} v_\phi \iiint d^3p\frac{\beta_z-\beta_\phi}{\beta_\phi-\beta_z}\frac{\partial f_s^{(0)}}{\partial p_\tau} \nonumber\\
  +\sum_s \frac{\omega_{ps}^2 }{\omega^2} \beta_\phi^2 \iiint d^3p\frac{1}{\beta_\phi-\beta_z}\frac{\partial f_s^{(0)}}{\partial p_\tau} \, .\label{eqa:2}
\end{align}
The first integral of Eq. \eqref{eqa:2} being odd and of the Cauchy type, its contribution vanishes. Moreover,  Eq. \eqref{eq:mj} leads to 
\begin{equation}\label{eq:dmj}
  \frac{\partial f_s^{(0)}}{\partial p_\tau}=-\mu_s(\beta_\tau-\beta_{ds\tau})f_s^{(0)} \, .
\end{equation}
Combining the above equations yields
\begin{equation}\label{eqa:3}
\epsilon_{zz}=1+\sum_s \frac{\omega_{ps}^2\mu_s}{k_z^2c^2}
-\sum_s \frac{\omega_{ps}^2\mu_s}{k_z^2c^2}(\beta_\phi-\beta_{ds})\iiint d^3p \frac{f_s^{(0)}}{\beta_\phi-\beta_z} \,.
\end{equation} 
This triple integral can be reduced to a much more tractable one-dimensional quadrature by changing to velocity variables in cylindrical coordinates
along the wave vector $\mathbf{v}=(v_{\perp}\cos(\theta),v_{\perp}\sin(\theta),v_{\parallel})$:
\begin{equation}\label{eqa:4}
\epsilon_{zz}=1
-\sum_s \frac{\omega_{ps}^2\mu_s}{k^2c^2}(\beta_\phi-\beta_{ds})\tilde{B}(\beta_\phi)\, ,
\end{equation} 
with
\begin{align}
&\tilde{B}(\beta_\phi)=\int d\beta \frac{f_{\tilde{B}}}{v_\phi-v_z} \, , \label{eqa:5} \\
&f_{\tilde{B}}= \frac{\gamma^3 e^{-h_s}}{h_s^5} \Big[ (h_s+1 ) (2\rho_s^2+\nu_s^2) + \rho_s^2 h_s^2\Big]\, ,  \label{eqa:6} \\
&h_s(\alpha=0)=\mu_s\gamma(1-\beta_{ds}\beta) \,.
\end{align}

\section{Basic formulae for $\alpha=\pi/2$}\label{ap:alpha_pis2}

Substituting $\langle E_zE_z^*\rangle_{k_y,\omega}= (\omega/k_yc)^2 \langle B_xB_x^*\rangle_{k_y,\omega}$ into Eq. \eqref{eq:fluctuations}
leads to
\begin{align}\label{eq:fluctubx05pi0}
\langle B_xB_x^*\rangle_{k_y,\omega}& = k_y^2\frac{\vert \epsilon_{yy}\vert^2 \langle j_z j_z^*\rangle_{\mathbf{k},\omega}}
{\vert(\omega^2 \epsilon_{zz}-k_y^2)\epsilon_{yy}-\omega^2 \epsilon_{yz}^2 \vert^2} \nonumber\\ 
+&k_y^2\frac{\vert\epsilon_{yz}\vert^2 \langle j_y j_y^* \rangle_{k_y,\omega}
+2\Re{(\epsilon_{yy}\epsilon_{yz}^*)} \langle j_yj_z^* \rangle_{k_y,\omega}}
{\vert(\omega^2 \epsilon_{zz}-k_y^2)\epsilon_{yy}-\omega^2 \epsilon_{yz}^2 \vert^2} \, .
\end{align} 
The susceptibility tensor is given by Eqs. \eqref{eq:esilonxx}-\eqref{eq:esilonyz} with $\alpha=\pi/2$:
\begin{eqnarray}
\epsilon_{xx}&=&1-\sum_{s}\frac{2\pi F_s \mu_{s} \omega_{ps}^2}{\omega k_y} D_s \label{eq:epsilonp05ixx}\\
\epsilon_{yy}&=&1-\sum_{s}\frac{2\pi F_s \mu_{s}\omega_{ps}^2}{\omega k_y} B_s\label{eq:epsilonp05iyy} \\
\epsilon_{zz}&=&1-\sum_{s}\frac{2\pi F_s \mu_{s}\omega_{ps}^2}{\omega k_y} A_s + \sum_{s}\frac{\mu_{s}\omega_{ps}^2}{\omega^2}\beta^2_{ds}\label{eq:psilonp05izz} \\
\epsilon_{yz}&=&\sum_{s}\frac{2\pi F_s \mu_{s}\omega_{ps}^2}{\omega k_y} C_s \label{eq:epsilonp05iyz}.
\end{eqnarray}
The corresponding sources are
\begin{eqnarray} 
\langle j_xj_x^{*}\rangle_{k_y,\omega}=H(1-|\beta_\phi|)\sum_{s}\frac{(2\pi)^2 F_s \mu_{s} n_s q_s^2}{\omega k_y} f_{D} \,, \label{eq:xxsourcespis2} \\
\langle j_yj_y^{*}\rangle_{k_y,\omega}= H(1-|\beta_\phi|)\sum_{s}\frac{(2\pi)^2 F_s \mu_{s}n_sq_s^2}{\omega k_y}f_{B} \, ,\label{eq:yysourcespis2} \\
\langle j_zj_z^{*}\rangle_{k_y,\omega}= H(1-|\beta_\phi|)\sum_{s}\frac{(2\pi)^2 F_s \mu_{s}n_sq_s^2}{\omega k_y} f_{A} \, , \label{eq:zzsourcespis2} \\
\langle j_yj_z^{*}\rangle_{k_y,\omega}= H(1-|\beta_\phi|)\sum_{s}\frac{(2\pi)^2 F_s \mu_{s}n_sq_s^2}{\omega k_y} f_{C} \,.\label{eq:yzsourcespis2}
\end{eqnarray}

\section{Calculation of the susceptibility tensor $\boldsymbol{\epsilon}(k_y,\omega = 0)$ for $\alpha=\pi/2$}
\label{sec:w0_mag_spectrum}

Equation \eqref{eq:tenseur_eps} with $\alpha =\pi/2$ yields
\begin{align}
\epsilon_{yy}&=1-\sum_s \frac{\omega_{ps}^2\mu_s}{\omega} \int_{\mathbb{R}} d^3p \frac{v_y v_y}{\omega - k_y v_y}f^{(0)}_s  \,, \label{eq:1yy}\\
\omega \epsilon_{yz}&=-\sum_s\omega_{ps}^2\mu_s \int_{\mathbb{R}} d^3p \frac{v_y v_z}{\omega - k_yv_y}f^{(0)}_s \,,  \label{eq:1yz}\\
\omega^2 \epsilon_{zz}&=\omega^2-\sum_s\omega_{ps}^2\mu_s\omega \int_{\mathbb{R}} d^3p \frac{v_z v_z}{\omega - k_y v_y}f^{(0)}_s \,.  \nonumber\\
&+\sum_s\omega_{ps}^2\mu_s\beta_{ds}^2\label{eq:1zz}
\end{align}
Taking $\omega\rightarrow 0$ in Eqs. \eqref{eq:1yy}-\eqref{eq:1zz} and using 
\begin{equation}
\lim_{v_\phi \rightarrow 0} \frac{1}{v_\phi}\int d^3p \frac{v_y^2 f_s^{(0)}}{v_\phi-v_y} = \int d^3p f_s^{(0)} = 1 \,,
\end{equation}
 gives 
\begin{align}
\lim_{\omega\rightarrow 0}&\epsilon_{yy}=1+\sum_s \frac{\omega_{ps}^2\mu_s}{k_y^2}  \,, \label{eqc:2yy}\\
\lim_{\omega\rightarrow 0}&\omega \epsilon_{yz}=-\sum_s\frac{\omega_{ps}^2\mu_s}{k_y}\beta_{ds}  \,,\label{eqc:2yz}\\
\lim_{\omega\rightarrow 0}&\omega^2 \epsilon_{zz}=\sum_s\omega_{ps}^2\mu_s\beta_{ds}^2 \,. \label{eqc:2zz}
\end{align}

\section{Calculation of the $\langle \mathbf{E} \mathbf{E}^\dagger \rangle_{k_y}$ and $\langle B_z B_z^*\rangle_{k_y}$ spectra for $\alpha=\pi/2$}
\label{sec:ExEx_EyEy_EzEz_BzBz}

In the case of an electron or pair plasma with equal temperatures and velocities, Eq. \eqref{eq:fluctuations} yields
\begin{align}\label{fluctuez05pi}
 &\langle E_z E_z^*\rangle _{k_y,\omega} =-\frac{2T}{\epsilon_0\omega}
 \Im{\left(\frac{\omega^2\epsilon_{yy}}{(\omega^2 \epsilon_{zz}-k_y^2c^2)\epsilon_{yy}-\omega^2 \epsilon_{yz}^2 }\right)} \,. \\
 &\langle E_x E_x^*\rangle_{k_y,\omega} =-\frac{2T}{\epsilon_0 (\omega - k_y v_{d})} \Im{\left(\frac{1}{\epsilon_{xx}-\frac{k^2c^2}{\omega^2}}\right)}.
\end{align} 
Proceeding as in Sec.\ref{sec:long_k_spec}, we obtain the total and supraluminal spatial $\langle E_z E_z^* \rangle_{k_y}$ spectra:
\begin{align}\label{int_ez2_pis2}
  &\langle E_z E_z^* \rangle _{k_y} =\frac{T}{\epsilon_0}\, , \\
  &\langle E_z E_z^* \rangle^{|\beta_\phi|>1}_{k_y} = \frac{T}{\epsilon_0} \sum_{\omega = \omega_{1,2S}}\frac{\omega\epsilon_{yy}}{\partial D/\partial \omega}.
\end{align} 
Likewise, one can readily derive the total and supraluminal spatial $\langle E_x E_x^* \rangle_{k_y}$ and $\langle B_z B_z^* \rangle_{k_y}$ spectra
\begin{align}\label{eq:int_transv2}
&\epsilon_0\langle E_x E_x^*\rangle_{k_y} = \frac{1}{\mu_0}\langle B_z B_z^*\rangle_{k_y} = T\, , \\
&\langle E_x E_x^*\rangle^{|\beta_\phi|>1}_{k_y} =-\sum_{\omega=\omega_{XS}}\frac{T}{\epsilon_0(\omega - k_yv_{d})} \frac{1}{\partial G/\partial \omega} \,,
\label{eq:int_transv_supralume_pis2}\\
&\langle B_z B_z^*\rangle^{|\beta_\phi|>1}_{k_y} =-\sum_{\omega=\omega_{XS}}\frac{T}{\epsilon_0 (\omega - k_yv_{d})} \frac{1}{v_{\phi}^2
\partial G/\partial \omega}  \label{eq:int_transv_supralumb_pis2} \,.
\end{align}
Here $\omega_{XS}(k_y)$ denotes the supraluminal solution of the dispersion relation Eq. \eqref{eq:dispe1}. The latter is recast in the form
\begin{equation}\label{supralumxx2}
  k_y^2c^2 = \frac{1}{\beta_{\phi}^2-1}\sum_s 2\pi F_s \mu_s \omega_{ps}^2 (\beta_{\phi}-\beta_{ds})
  D_s(\beta_\phi) \,.
\end{equation}

Finally, combining  Eqs. \eqref{eq:fluctuations} and \eqref{eq:epsilonp05ixx}-\eqref{eq:yzsourcespis2} yields
\begin{align}\label{fluctuey05pi0}
\langle E_yE_y^*\rangle_{k_y,\omega} &=\frac{\omega^2}{\vert D \vert^2}
\Big[ \vert \epsilon_{yz} \vert^2\langle j_z j_z^{*} \rangle_{k_y,\omega}  \nonumber \\   
&+ \vert\omega^2\epsilon_{zz}-k_y^2 \vert^2 \langle j_y j_y^{*}\rangle_{k_y,\omega} \nonumber\\ 
&+ 2\Re{\left((\omega^2\epsilon_{zz}-k_y^2c^2)\epsilon_{yz}^* \right)}\langle j_yj_z^{*}\rangle_{k_y,\omega} \Big] \, .
\end{align}
For an electron/pair plasma with equal temperatures and velocities, this equation reduces to
\begin{equation}\label{fluctuey05pi}
\langle E_yE_y^*\rangle_{k_y,\omega} =-\frac{2T}{\epsilon_0\omega}
\Im{\left(\frac{\omega^2\epsilon_{zz}-k_y^2c^2}{(\omega^2 \epsilon_{zz}-k_y^2c^2)\epsilon_{yy}-\omega^2 \epsilon_{yz}^2 }\right)} \,.
\end{equation} 
There follow the spatial spectra
\begin{align}
&\langle E_y E_y^*\rangle_{k_y} =\frac{T}{\epsilon_0} \frac{\omega_p^2}{k_y^2c^2+\omega_p^2 \mu/\gamma_d^2}\, , \label{int_ey2_pis2} \\
&\langle E_y E_y^*\rangle^{|\beta_\phi|>1}_{k_y} = \frac{T}{\epsilon_0}\sum_{\omega = \omega_{1,2S}}
\frac{\omega^2\epsilon_{zz}-k_y^2c^2}{\omega \partial D/\partial \omega} \, .
\label{int_ey2_pis2_suparlum}
\end{align} 

%\bibliography{biblio}

\end{document}